\documentclass[10pt]{iopart}

\usepackage{amssymb}
\usepackage{graphicx}

\usepackage{hyperref}

\usepackage{mathptmx}

\newcommand{\Q}{\mathsf{Q}}
\newcommand{\V}{\mathsf{V}}
\newcommand{\Pp}{\mathsf{p}}
\newcommand{\PP}{\mathsf{P}}
\renewcommand{\PM}{\mathsf{M}}
\newcommand{\Pm}{\mathsf{m}}
\newcommand{\D}{\mathsf{J}}
\newcommand{\B}{\mathsf{B}}
\newcommand{\R}{\mathsf{R}}
\newcommand{\T}{\mathsf{T}}
\newcommand{\RT}{\mathsf{S}}
\newcommand{\ps}{P_\mathrm{s}}
\newcommand{\pa}{P_\mathrm{a}}

\newcommand{\mm}[2]{\Pm^{(#1)}_{#2}}

\newcommand{\psf}{P_{\mathrm{sf}}}
\newcommand{\psb}{P_{\mathrm{sb}}}
\newcommand{\pbs}{P_{\mathrm{bs}}}
\newcommand{\pss}{P_\mathrm{ss}}

\newcommand{\mn}[2]{\mathsf{m}^{(#1)}_{#2}}
\newcommand{\an}[2]{\mathsf{a}^{(#1)}_{#2}}
\newcommand{\bn}[2]{\mathsf{b}^{(#1)}_{#2}}
\newcommand{\cn}[2]{\mathsf{c}^{(#1)}_{#2}}
\newcommand{\dn}[2]{\mathsf{d}^{(#1)}_{#2}}
\newcommand{\An}[3]{\mathsf{A}^{(#1)}_{#2}(#3)}
\newcommand{\Bn}[3]{\mathsf{B}^{(#1)}_{#2}(#3)}
\newcommand{\Cn}[3]{\mathsf{C}^{(#1)}_{#2}(#3)}
\newcommand{\Dn}[3]{\mathsf{D}^{(#1)}_{#2}(#3)}

\newcommand{\rr}{\bi{r}}
\newcommand{\vv}{\bi{v}}
\newcommand{\ee}{\bi{e}}

\newcommand{\defeq}{\equiv}
\newcommand{\eqdef}{\equiv}

\newcommand{\ii}{i}
\newcommand{\onehalf}{{_1\over^2}}
\newcommand{\ds}{\scriptstyle}

\eqnobysec

\begin{document}
\title[Diffusion coefficients for persistent random walks]
{Diffusion coefficients for multi-step persistent random walks on lattices}  
\author{Thomas Gilbert\dag, David P. Sanders\ddag}
\address{\dag Center for Nonlinear Phenomena and Complex Systems, Universit\'e
  Libre  de Bruxelles, C.~P.~231, Campus Plaine, B-1050 Brussels, Belgium\\
  \ddag Departamento de F\'isica, Facultad de Ciencias, Universidad Nacional
  Aut\'onoma de M\'exico, 04510 M\'exico D.F., Mexico} 

\ead{thomas.gilbert@ulb.ac.be, dps@fciencias.unam.mx}

\begin{abstract}
We calculate the diffusion coefficients of persistent random walks on lattices,
where the direction of a walker at a given step depends on the memory of a
certain number of previous steps. In particular, we describe a simple
method which enables us to obtain explicit expressions for the diffusion
coefficients of walks with two-step memory on different classes of one-,
two- and higher-dimensional lattices. 
\end{abstract}

\submitto{J.~Phys.~A: Math.~Theor.}


\section{\label{sec.int}Introduction}

Random walks are widely used throughout physics as a model for systems in which
the state of the system can be viewed as evolving in a stochastic way from
one time step to the next.  Their properties have been extensively
explored and the techniques to study them are well developed \cite{We94}.
In particular, at a large scale, random walks behave diffusively.

The most-studied case is that of random walkers which have no memory of their
past history. Many physical applications, however, call for a model in which
the choice of possible directions for the walker's next 
step are given by probabilities which are influenced by the path it took prior
to
making that choice, so that its jumps are correlated; this is often
called a \emph{persistent} or \emph{correlated} random walk \cite{HK87}.

A walk with zero memory corresponds to the
Bernoulli process of the usual uncorrelated random walk. Walks with single-step
memories are the most commonly studied cases of
persistent random walks, where the random walker determines the
direction it takes at a given step in terms of the direction taken on the
immediately preceding step \cite{We94}. Such walks -- not necessarily
restricted to lattices, as will be the case here -- were first discussed in
the context of Brownian motion \cite{Fu20} and fluid dynamics \cite{Ta22},
and have 
since found many applications in the physics literature, most prominently
in polymer conformation theory \cite{Ku36} and tracer diffusion in metals
\cite{Ma59}, but also in relation to the telegrapher's equation in the
context of thermodynamics \cite{We02}. Previous works dealing with random
walks on lattices with higher-order memory effects include that of
Montroll \cite{Montroll1950}, with applications to models of polymers, and
Bender and Richmond \cite{Bender1984}.  

The state of the walker is thus specified by two variables,
its location and the direction it took at the preceding step. The
statistical properties of such persistent walks can be described by simple
Markov chains and have already been thoroughly investigated in the
literature; see in particular ref.~\cite{RH81}. We will only provide a short
review of results relevant to our purposes, with specific emphasis on the
diffusive properties.

The statistics of random walks with 
multi-step memory can in principle be analysed in terms of Markov
chains, in a similar way to their
single-step memory counterpart.
However, the number of states of these chains grows exponentially with 
the number of steps accounted for. This is the source of great technical
difficulties, which are present already at the level of two-step processes. 

Of specific interest to us are random walks with two-step memory. Among
the
class of persistent walks under consideration, these are the simplest case
beyond those with single-step memory, and are therefore relevant to
problems dealing with the persistence of motion of tracer particles where
the single-step-memory approximation breaks down. An example where this
occurs is given 
in recent work by the present authors,
on diffusion in a class of periodic billiard tables \cite{GilbertSanders09}.

The paper is organised as follows. The general framework of walks on
lattices is briefly reviewed in section \ref{sec.dif}, where we provide the
expression of the diffusion coefficient of such walks in terms of the
velocity auto-correlations. Successive approximation schemes for the
computation of these auto-correlation functions are presented in
sections \ref{sec.nma}, \ref{sec.1sma}, and \ref{sec.2sma}, pertaining to
the number of steps of memory of the walkers, respectively 0, 1, and
2. Specific examples are discussed, namely the one-dimensional lattice and
the two-dimensional square, honeycomb and triangular lattices,
and their diffusion coefficients are computed.  Some of the details of the
computations presented in section \ref{sec.2sma} are deferred to appendices
\ref{app.hc} and \ref{app.sq}. Section \ref{sec.2rev} provides an
alternative derivation of the diffusion coefficients of two-step memory
persistent walks with special left--right symmetries. Conclusions are drawn
in section \ref{sec.con}.

\section{\label{sec.dif}Diffusion on a lattice}

We consider the motion of independent tracer particles undergoing random
walks on a regular lattice $\mathcal{L}$. Their trajectories are specified
by their initial position $\rr_0$ at time $t=0$, and the sequence
$\{\vv_{0},\ldots, \vv_{n}\}$ of the successive values  $\vv_{i}\in
\mathcal{V}_{\rr_{i}}$ of their direction vectors at positions
$\rr_{i}$, where $\mathcal{V}_{\rr_{i}}$ denotes the space of
direction vectors allowed at site $\rr_i$, which point to the lattice sites
adjacent to $\rr_i$. Here we consider dynamics in discrete time, so that the
time sequences are
simply assumed to be incremented by identical time steps $\tau$ as the
tracers move from site to site. In the sequel we will loosely refer to the
direction vectors as velocity vectors; they are in fact dimensionless unit 
vectors.

Examples of such motions are random walks on one- and two-dimensional
lattices such as honeycomb, square and triangular lattices, but also
include persistent 
random walks where memory effects must be accounted for, i.e.\ when
the probability of occurrence of $\vv_n$ depends on the past history 
$\vv_{n-1}, \vv_{n-2},\ldots$.

The quantity we will be concerned with is the diffusion coefficient $D$ of
such persistent processes, which measures the linear growth in time of the
mean-squared displacement of walkers. This can be written in terms of velocity
auto-correlations using the Taylor--Green--Kubo expression:
\begin{equation}
  D = \frac{\ell^2}{2 d \tau} \left[1 + 2 \lim_{K\to\infty}
    \sum_{n = 1}^K \langle
    \vv_0 \cdot \vv_{n} \rangle \right]\,,
  \label{dcoef}
\end{equation}
where $d$ denotes the dimensionality of the lattice $\mathcal{L}$, and $\ell$
is the lattice spacing. The (dimensionless) velocity auto-correlations
are computed as averages
$\langle \cdot \rangle$ over the 
equilibrium distribution $\mu$, so that the problem reduces to computing
\begin{equation}
  \langle \vv_0 \cdot \vv_{n} \rangle
  = \sum_{\vv_0, \ldots, \vv_{n}}  \vv_0 \cdot \vv_{n} \, 
  \mu(\{\vv_{0}, \ldots, \vv_{n}\}) \, .
  \label{v0vn}
\end{equation}

As reviewed below, this can be easily carried out in the simple examples
of random walks with zero- and single-step memories. The main achievement
of this paper is to describe the computation of the velocity
auto-correlations of random walks with two-step memory.  
All these cases involve factorisations of the measure $\mu(\{\vv_{0},
\ldots, \vv_{n}\})$ by products of probability measures 
which depend on a number of velocity vectors, equal to the number of steps
of memory of the walkers. These measures will be denoted by $p$ throughout
the paper. 

The next three sections,  sections~\ref{sec.nma}, \ref{sec.1sma} and
\ref{sec.2sma},
are devoted to the computation of the diffusion coefficient (\ref{dcoef})
for random walks with zero-, one- and two-step memories, respectively. 

\section{\label{sec.nma}No-Memory Approximation (NMA)}

In the simplest case, the walkers have no memory of their
history as they proceed to their next position. This gives a Bernoulli process
for the velocity trials, for which the probability measure factorises:
\begin{equation}
  \mu(\{\vv_{0},\dots, \vv_{n}\}) = \prod_{i = 0}^n
  p(\vv_{i})\,.
\end{equation}
Given that the lattice is isotropic and that $p$ is uniform, the velocity
auto-correlation (\ref{v0vn}) vanishes:
\begin{equation}
  \langle \vv_{0} \cdot \vv_{n} \rangle = \delta_{n,0}\,.
\end{equation}
The diffusion coefficient of the random walk without memory is then given
by 
\begin{equation}
  D_\mathrm{NMA} = \frac{\ell^2}{2d\tau}\,.
\end{equation}

\section{\label{sec.1sma}One-Step Memory Approximation (1-SMA)}

We now assume that the velocity vectors obey a Markov process for which
$\vv_{n}$ takes on different values according to the
velocity at the previous step $\vv_{n-1}$.  We may then write
\begin{equation}
  \mu(\{\vv_{0},\dots, \vv_{n}\}) = \prod_{i = 1}^n 
  P(\vv_{i}|\vv_{i-1}) p(\vv_{0})\,.
\end{equation}
Here, $P(\mathbf{b} | \mathbf{a})$ denotes the $1$-step conditional probability
that the walker moves in a direction $\mathbf{b}$,
given that
it had direction $\mathbf{a}$ at the previous step.

We denote by $z$ the coordination number of the lattice, i.e.\ the number of
neighbouring sites accessible from each site,  and we denote by $\R$ the
rotation operation which takes a vector $\vv$ through all the lattice
directions $\vv, \R\vv,\dots,\R^{z-1}\vv$. In general,
the set of allowed orientations of $\vv$ depends on the lattice site,
such as in the two-dimensional honeycomb lattice. We denote by $\T$
the symmetry operator that maps a cell to its neighbors, which corresponds
simply to the identity for square lattices and to a reflection for the
honeycomb lattice. 

The idea of our calculation is to express each velocity vector $\vv_k$ in
terms of the first one, $\vv_0$, as $\vv_k = \R^{i_k}\T^k \vv_0$,
where $i_k$ lies between $0$ and $z-1$.
Substituting this into the expression for the velocity auto-correlation   
$\langle \vv_{0} \cdot \vv_{n} \rangle$, equation (\ref{v0vn}), we obtain 
\begin{equation}
  \fl \sum_{\vv_{0},\ldots, \vv_{n}}
  \vv_{0} \cdot \vv_{n}   \prod_{i = 1}^n 
  P(\vv_{i}|\vv_{i-1}) p(\vv_{0})
  =  \sum_{i_0,\ldots,i_n = 1}^z
  \vv_{0} \cdot \R^{i_n}\T^n \vv_0\,
  \Pm_{i_n,i_{n-1}} \cdots \Pm_{i_1,i_0} \Pp_{i_0}\,,
  \label{v0vn0}
\end{equation}
where
\begin{equation}
  \Pm_{i_n,i_{n-1}} \defeq P(\R^{i_n}\T^n \vv_0|\R^{i_{n-1}}\T^{n-1} \vv_0) 
  \label{def-M}
\end{equation}
are the elements of the stochastic matrix $\PM$ of the Markov chain
associated to the persistent 
random walk, and $\Pp_i \defeq p(\mathbf{i})$ are the elements of its invariant
(equilibrium)
distribution, denoted $\PP$, evaluated with a
velocity in the $i$th lattice direction. The invariance of $\PP$ is expressed as
$\sum_{j}\Pm_{i,j}\Pp_j = \Pp_i$. These
notations will be used throughout this article.

The terms involving $\PM$ in (\ref{v0vn0}) constitute the matrix product of
$n$ copies of $\PM$. Furthermore, 
since the invariant distribution is uniform over the $z$ possible lattice
directions,  we can choose an arbitrary direction for $\vv_0$, and hence
write 
\begin{equation}
  \langle \vv_0 \cdot  \vv_n \rangle
  = \vv_0 \cdot \T^n\vv_0
    \mm{n}{1,1} + \vv_0 \cdot \R\T^n\vv_0 \mm{n}{2,1} + \cdots
    + \vv_0 \cdot \R^{z-1}\T^n\vv_0 \mm{n}{z,1}\,,
  \label{v0vn1sma}
\end{equation}
where $\mm{n}{i,j}$ denote the elements of $\PM^n$.

Under special symmetry assumptions to be discussed in the examples below,
one has 
\begin{equation}
  \langle \vv_0 \cdot  \vv_n \rangle
  = \langle \cos \theta\rangle^n\,,
  \label{idthetan}
\end{equation}
where $\langle \cos \theta\rangle$ denotes the average angle between two
successive velocity vectors.
It is then a general, well-known, result for such symmetric persistent
random walks with
single-step memory \cite{HK87} that their diffusion coefficients have the form
\begin{equation}
  D_\mathrm{1SMA} = D_\mathrm{NMA} 
  \frac{1 + \langle \cos \theta\rangle}
  {1 - \langle \cos \theta\rangle} 
  \qquad (\mathrm{symmetric\, walks}).
    \label{dcoef1spa}
\end{equation}

\begin{figure}[htb]
  \centering
  (a) \includegraphics[width=.3\textwidth]{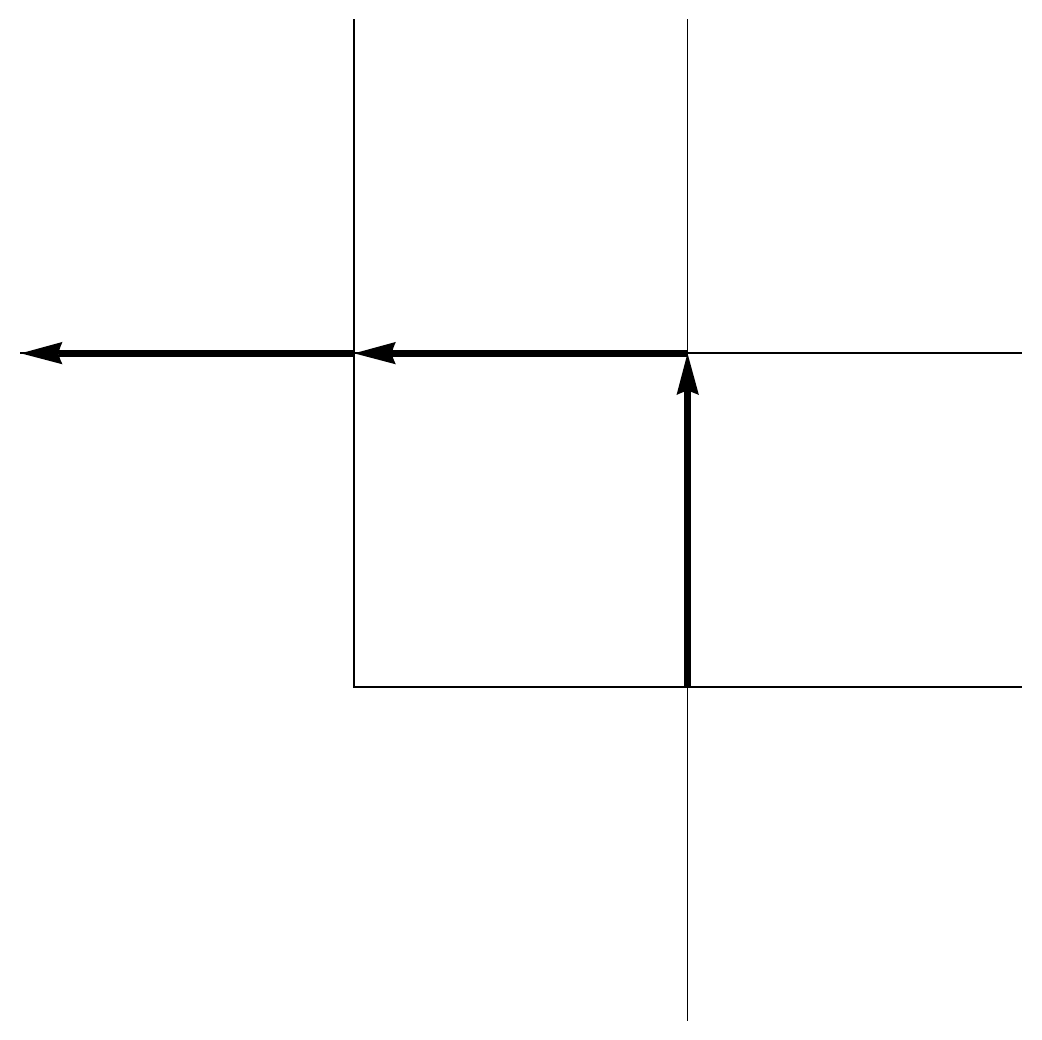}
  \qquad
  (b) \includegraphics[width=.3\textwidth,angle=90]{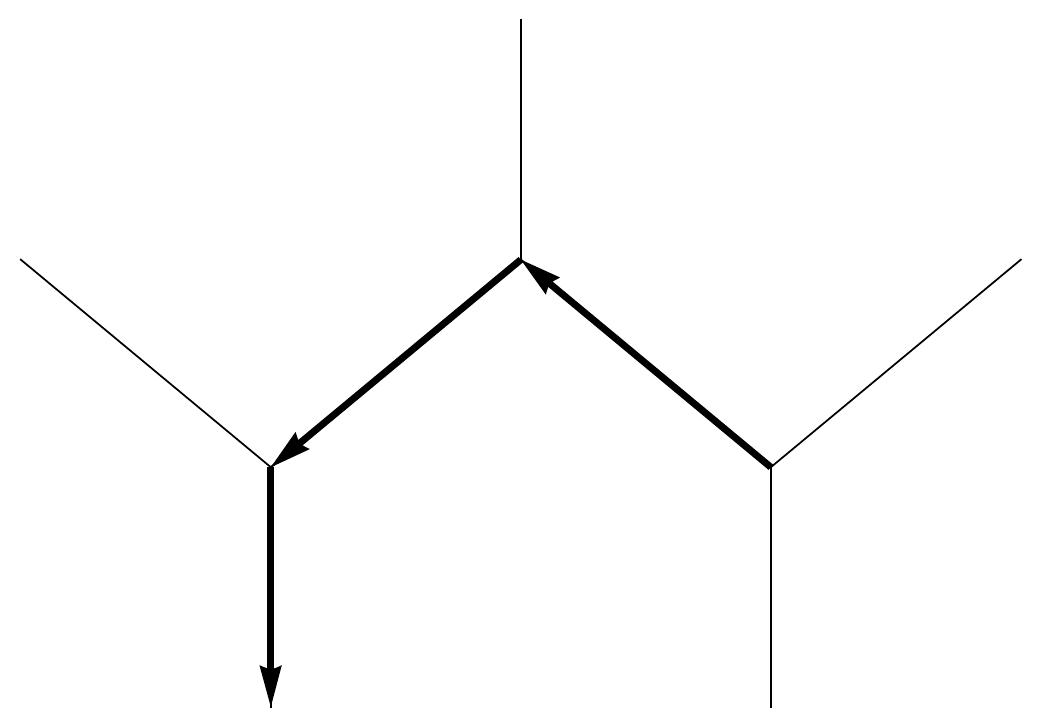}
  \caption{Examples of walks on (a) square and (b) honeycomb
    lattices. Note the inversion of the allowed directions at
    neighbouring sites on the honeycomb lattice.}
  \label{fig.walk}
\end{figure}

\begin{figure}[htb]
  \centering
  (a) \includegraphics[width=.3\textwidth,angle=0]{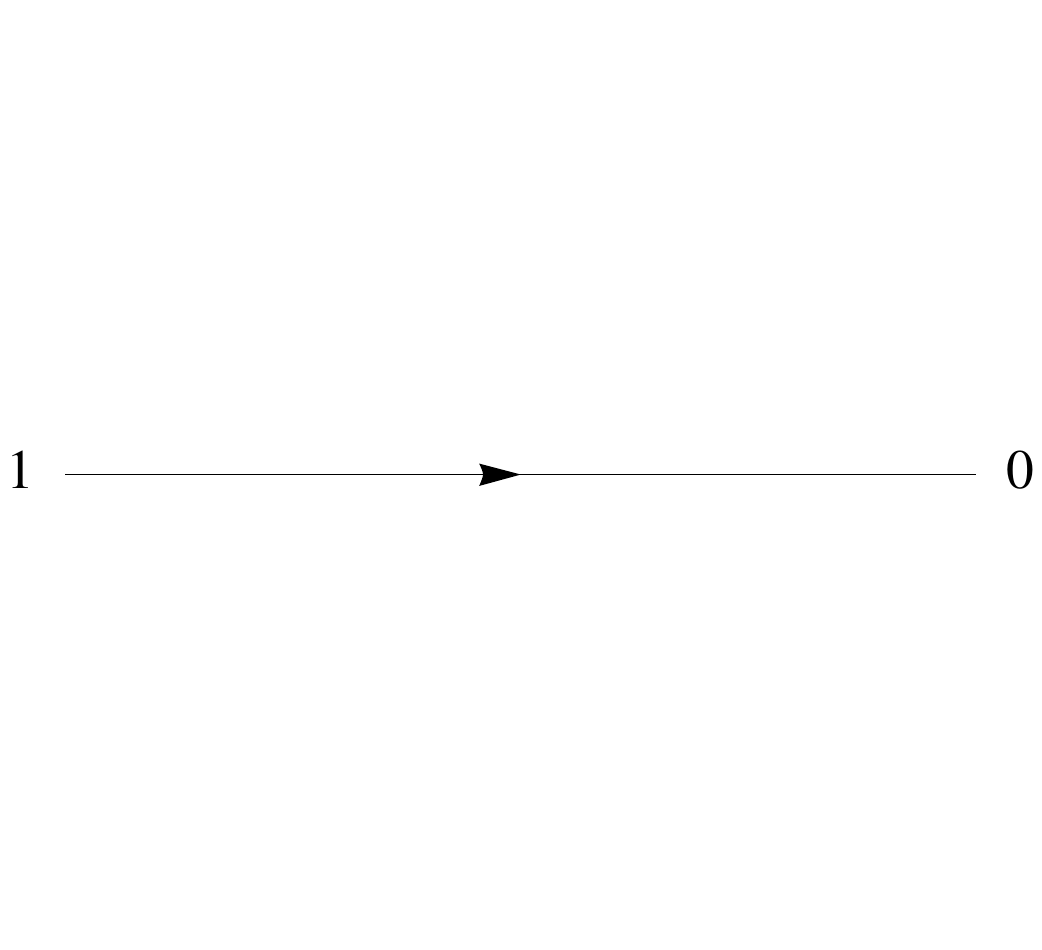}
  \qquad
  (b) \includegraphics[width=.3\textwidth,angle=0]{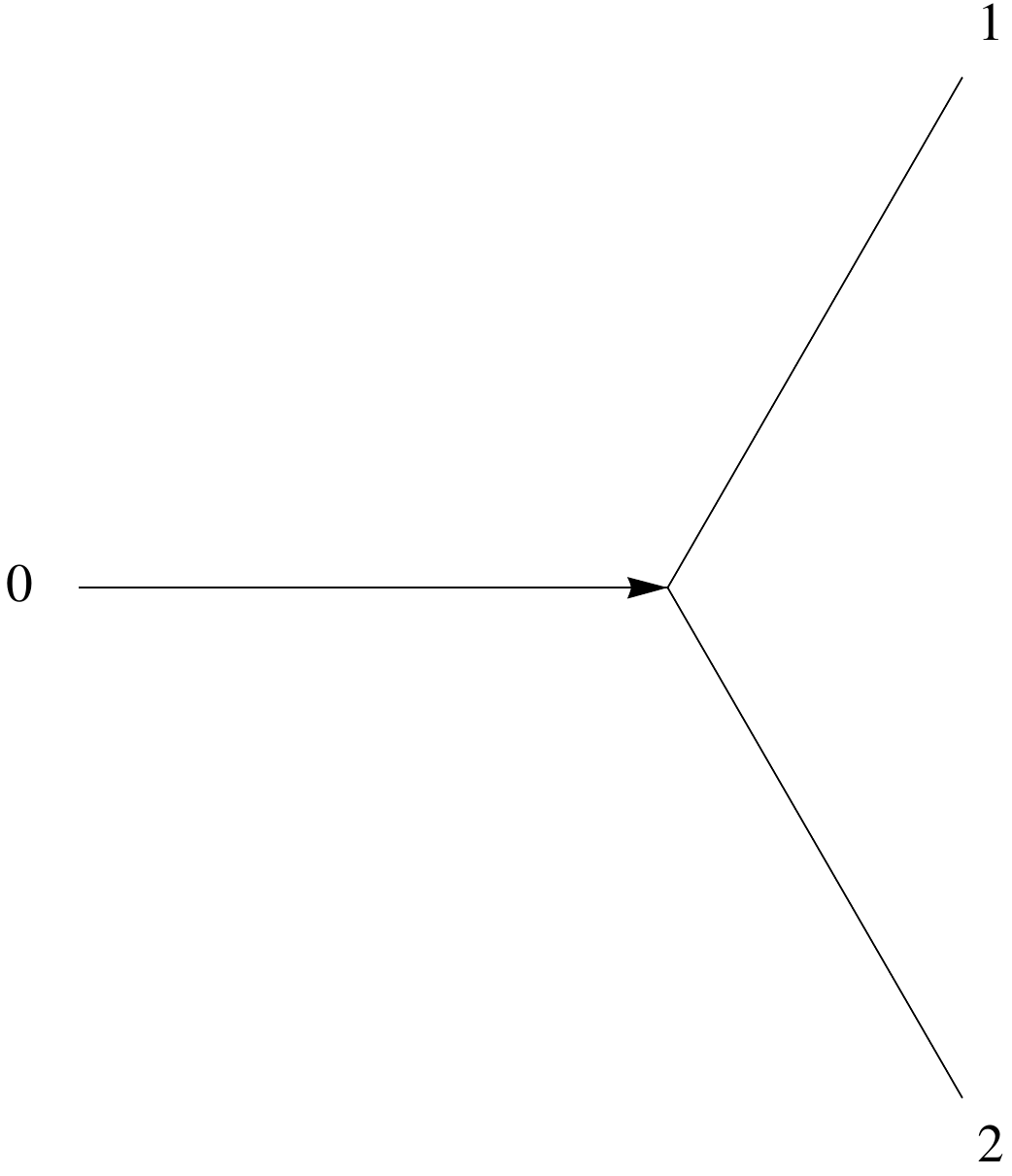}\\
  (c) \includegraphics[width=.3\textwidth,angle=0]{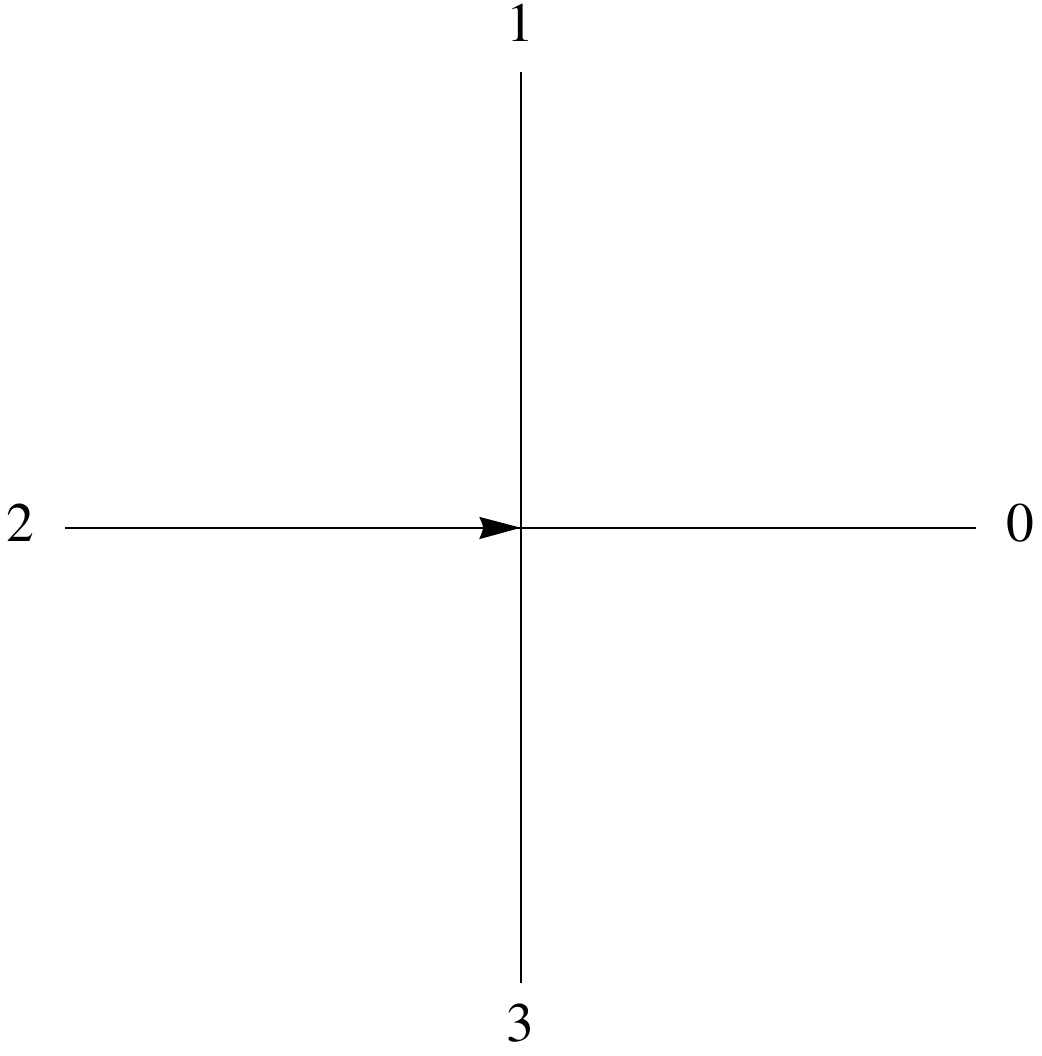}
  \qquad
  (d) \includegraphics[width=.3\textwidth,angle=0]{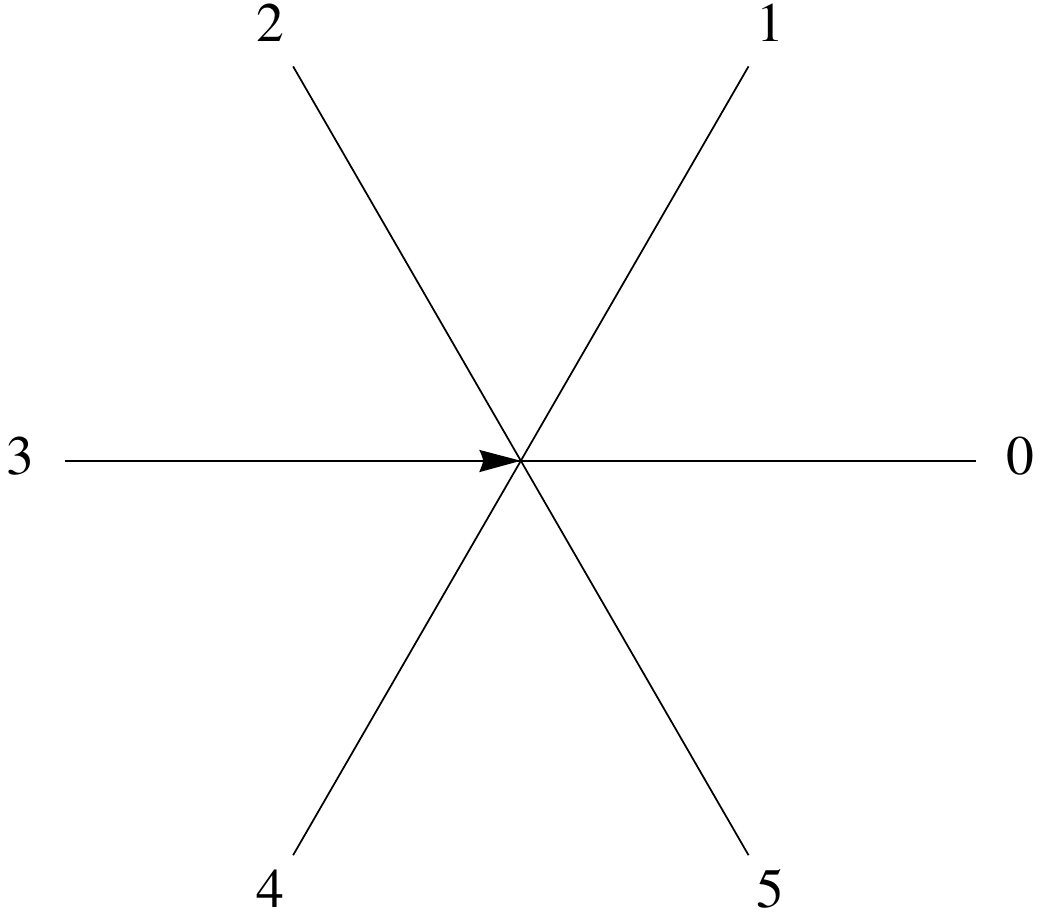}
  \caption{The possible directions of motion at a given step for different
lattices, \emph{relative} to the incoming direction which is shown by the arrow,
are labelled from $0$ to $z-1$,
    corresponding to the angle $2\pi j/z$ that the lattice direction $j$ makes
    with respect to the reference direction, up to a reflection inversion
    in the case of the honeycomb lattice. (a) One-dimensional lattice ($z =
2$);  (b) honeycomb lattice ($z=3$); (c) square lattice ($z = 4$);
    (d) triangular lattice ($z=6$) .}
  \label{fig.num}
\end{figure}

The actual value of the diffusion coefficient depends on the probabilities
$P(\R^j\T\vv|\vv)$, which are parameters of the model. Specific
applications of equation (\ref{dcoef1spa}) are given in the 
examples below, such as shown in figure \ref{fig.walk}. To simplify
the notation, we  denote the conditional probabilities of these walks by
$P_j$, where $j = 0,\dots, z-1$ corresponds to the relative angle $2\pi j/z$ of
the direction that the walker takes with respect to its previous step (up to a
reflection in the case of the honeycomb lattice). These conventions are
shown in figure \ref{fig.num}. 

\subsection{\label{sub.1sma1d} One-dimensional lattice}

The simplest case is that of a regular one-dimensional lattice. In this case,
each site is equivalent, and so $\T$ is the identity. Each velocity vector
$\vv_k$ has only two possible values, $v$ and $-v$, so that $\R$ is a
reflection. We denote by $P_0 \defeq P(v|v)$ the probability that
the random walker continues in the same direction at the next step, and by
$P_1 \defeq P(-v|v)$ the probability that it reverses direction.

It is easy to check that the velocity auto-correlation (\ref{v0vn1sma}) yields
(\ref{idthetan}): 
\begin{equation}
  \langle \vv_0\cdot\vv_n \rangle
  = (P_0 - P_1)^n = \langle \cos \theta \rangle^n\, ,
\end{equation}
and so the diffusion coefficient is given by
\begin{equation}
  D_\mathrm{1SMA} = D_\mathrm{NMA}\frac{P_0}{P_1} \,.
    \label{dcoef1spa1d}
\end{equation}

\subsection{\label{sub.1smasq}Two-dimensional square lattice}

On a two-dimensional square lattice, $\vv_k$ can take four possible values,
and each lattice site is again equivalent. Thus $\T$ is the identity and
$\R$ can be taken as an anticlockwise rotation by angle $\pi/2$. We
denote by $P_0 \defeq  P(\vv|\vv)$ the probability that the particle
proceeds in the same direction as on its previous step, by $P_1 \defeq P(\R
\vv|\vv)$ the probability that the particle turns to the left relative to
its previous direction, by $P_2 \defeq P(\R^2 \vv|\vv)$ the probability
that it  turns around, and by $P_3 \defeq P(\R^3 \vv|\vv)$ the probability
that it turns right.

From  (\ref{v0vn1sma}),  the velocity auto-correlation $\langle
\vv_0\cdot\vv_n \rangle$ is given by
\begin{equation}
  \langle \vv_0\cdot\vv_n \rangle = \mm{n}{1,1} - \mm{n}{1,3}\,.
  \label{v0vn1smasqexp}
\end{equation}
The transition matrix $\PM$ given by 
(\ref{def-M})  is thus the following cyclic matrix:
\begin{equation}
  \PM = 
  \left(
    \begin{array}{c@{\enspace}c@{\enspace}c@{\enspace}c}
      P_0 & P_1 & P_2 & P_3 \\
      P_3 & P_0 & P_1 & P_2 \\
      P_2 & P_3 & P_0 & P_1 \\
      P_1 & P_2 & P_3 & P_0
    \end{array}
  \right)\,.
\end{equation}

To calculate the elements $\mm{n}{i,j}$ of the powers $\PM^n$, it is possible to
compute the eigenvalues and eigenvectors of
$\PM$ and then decompose it as $\PM = \mathsf{Q} \cdot
\mathsf{L} \cdot \mathsf{Q}^{-1}$, where $\mathsf{L}$ is the diagonal
matrix of the eigenvalues of $\PM$ and $\mathsf{Q}$ the matrix of its
eigenvectors.
This procedure is, however, not necessary here, since we only
require the combination of the $\mm{n}{i,j}$ which appears in 
(\ref{v0vn1smasqexp}). To proceeed, we label the \emph{distinct} entries of
$\PM^n$ as $\an{n}{i}$, using the fact that $\PM^n$ is also cyclic if $\PM$ is:
\begin{equation}
  \PM^n = 
  \left(
    \begin{array}{c@{\enspace}c@{\enspace}c@{\enspace}c}
      \an{n}{1} & \an{n}{2}& \an{n}{3} & \an{n}{4}\\
      \an{n}{4} & \an{n}{1}& \an{n}{2} & \an{n}{3}\\
      \an{n}{3} & \an{n}{4}& \an{n}{1} & \an{n}{2}\\
      \an{n}{2} & \an{n}{3}& \an{n}{4} & \an{n}{1}
    \end{array}
  \right)\,.
  \label{mnasq}
\end{equation}
Writing $\PM^n = \PM \PM^{n-1}$, we can exploit the particular structure of
the matrix to reduce it from a $(4 \times 4)$-matrix to a $(2 \times
2)$-matrix, by considering the following differences:
\begin{eqnarray}
\fl
\left(
    \begin{array}{c}
      \an{n}{1} - \an{n}{3} \\
      \an{n}{2} - \an{n}{4} 
    \end{array}
  \right)
  &= 
  \left(
    \begin{array}{c@{\enspace}c}
      P_0 - P_2 & P_3 - P_1\\
      P_1 - P_3 & P_0 - P_2   
    \end{array}
  \right)
  \left(
    \begin{array}{c}
      \an{n-1}{1} - \an{n-1}{3} \\
      \an{n-1}{2} - \an{n-1}{4} 
    \end{array}
  \right)
  \nonumber\\
  &= 
  \left(
    \begin{array}{c@{\enspace}c}
      P_0 - P_2 & P_3 - P_1\\
      P_1 - P_3 & P_0 - P_2   
    \end{array}
  \right)^{n-1}
  \left(
    \begin{array}{c}
      P_0 - P_2 \\
      P_1 - P_3
    \end{array}
  \right).
\end{eqnarray}
The velocity auto-correlation (\ref{v0vn1smasqexp}) is thus given by
\begin{eqnarray}
  \eqalign
  \langle \vv_0\cdot\vv_n \rangle &= 
  \an{n}{1} - \an{n}{3} \nonumber\\ 
  &= \big(1 \enspace 0 \big)
  \left(
    \begin{array}{c@{\enspace}c}
      P_0 - P_2 & P_3 - P_1\\
      P_1 - P_3 & P_0 - P_2   
    \end{array}
  \right)^{n-1}
  \left(
    \begin{array}{c}
      P_0 - P_2 \\
      P_1 - P_3
    \end{array}
  \right).
  \label{v0vn1SMAsq} 
\end{eqnarray}

Summing the previous expression over all $n$, and using the fact that
$\sum_{n=0}^\infty
\mathsf{A}^n = (\mathsf{I} - \mathsf{A})^{-1}$ for a matrix $\mathsf{A}$, where
$\mathsf{I}$ is the identity matrix, we obtain the diffusion coefficient
(\ref{dcoef}) as
\begin{equation}
  \fl
  D_\mathrm{1SMA} = D_\mathrm{NMA}\left[1 +
    2 \Big(1\enspace 0 \Big)
    \left(
      \begin{array}{c@{\enspace}c}
        1 - P_0 + P_2 & P_1 - P_3\\
        P_3 - P_1 & 1 - P_0 + P_2   
      \end{array}
    \right)^{-1}
    \left(
      \begin{array}{c}
        P_0 - P_2 \\
        P_1 - P_3
      \end{array}
    \right)
  \right].
\end{equation}
For a symmetric process in which $P_1 = P_3$, which is
often imposed by a symmetry of the physical system, equation (\ref{idthetan})
holds, and the diffusion coefficient takes the form (\ref{dcoef1spa}), viz.
\begin{equation}
  D_\mathrm{1SMA} = D_\mathrm{NMA}\frac{1 + P_0 - P_2} {1 - P_0 + P_2}\,.
    \label{D1SMASq}
\end{equation}
If, however, $P_1 \ne P_3$, then equation (\ref{dcoef1spa}) is no longer
valid. Instead, we have the more complicated expression 
\begin{equation}
  D_\mathrm{1SMA} = D_\mathrm{NMA}
  \frac{1 - (P_0 - P_2)^2 - (P_1 - P_3)^2}
  {(1 - P_0 + P_2)^2 + (P_1 - P_3)^2}\,. 
\end{equation}
Such asymmetric walks are the lattice equivalent of the continuous-space
models of persistent random walks with chirality considered in \cite{La97}.

\subsection{\label{sub.1smahc}Two-dimensional honeycomb lattice} 

On the two-dimensional honeycomb lattice, shown in figure~\ref{fig.walk}(b),
each site has $z=3$ neighbours. Here, $\R$ is taken to be a
clockwise rotation\footnote{We take a clockwise rotation as opposed to an
  anticlockwise one in the other examples so that the lattice directions are
  still labelled anticlockwise.} by angle $2\pi/3$ and the arrangement
of neighbours differs by a reflection $\T$. We denote by $P_0 \defeq
P(-\vv|\vv)$ the probability that the particle turns around, $P_1 \defeq
P(-\R^2 \vv|\vv)$ that it turns left relative to its previous direction, and
$P_2 \defeq P(-\R \vv|\vv)$ that it turns right. The transition matrix $\PM$
is thus
\begin{equation}
  \PM = 
  \left(
    \begin{array}{c@{\enspace}c@{\enspace}c}
      P_0 & P_1 & P_2 \\
      P_2 & P_0 & P_1 \\
      P_1 & P_2 & P_0 \\
    \end{array}
  \right).
  \label{mhc}
\end{equation}

Proceeding as with the square lattice, we let
\begin{equation}
  \PM^n \equiv 
  \left(
    \begin{array}{c@{\enspace}c@{\enspace}c}
      \an{n}{1} & \an{n}{2} & \an{n}{3} \\
      \an{n}{3} & \an{n}{1} & \an{n}{2}\\
      \an{n}{2} & \an{n}{3} & \an{n}{1}
    \end{array}
  \right),
  \label{mnahc}
\end{equation}
and obtain the matrix equation
\begin{eqnarray}
  \fl
  \left(
    \begin{array}{c}
      - \an{n}{1} + \frac{1}{2} [\an{n}{2} + \an{n}{3}] \\
      - \an{n}{2} + \frac{1}{2} [\an{n}{1} + \an{n}{3}] \\
      - \an{n}{3} + \frac{1}{2} [\an{n}{1} + \an{n}{2}] \\
    \end{array}
  \right)
  = 
  \left(
    \begin{array}{c@{\enspace}c@{\enspace}c}
      P_0 & P_1 & P_2 \\
      P_2 & P_0 & P_1 \\
      P_1 & P_2 & P_0 \\
    \end{array}
  \right)
  \left(
    \begin{array}{c}
      - \an{n-1}{1} + \frac{1}{2} [\an{n-1}{2} + \an{n-1}{3}] \\
      - \an{n-1}{2} + \frac{1}{2} [\an{n-1}{1} + \an{n-1}{3}] \\
      - \an{n-1}{3} + \frac{1}{2} [\an{n-1}{1} + \an{n-1}{2}] \\
    \end{array}
  \right),
  \nonumber\\
  = \frac{1}{2}
  \left(
    \begin{array}{c@{\enspace}c@{\enspace}c}
      P_0 & P_1 & P_2 \\
      P_2 & P_0 & P_1 \\
      P_1 & P_2 & P_0 \\
    \end{array}
  \right)^{n-1}
  \left(
    \begin{array}{c}
      1 - 3 P_0 \\
      1 - 3 P_2 \\
      1 - 3 P_1 \\
    \end{array}
  \right).
  \label{eqmatrix}
\end{eqnarray}
The velocity auto-correlation (\ref{v0vn1sma}) is thus
\begin{eqnarray}
  \langle \vv_0\cdot\vv_n \rangle
  &=&       - \an{n}{1} + \frac{1}{2} [\an{n}{2} + \an{n}{3}] \,,
  \nonumber\\
  &=& \frac{1}{2} \Big(1\enspace0\enspace0\Big)
   \left(
    \begin{array}{c@{\enspace}c@{\enspace}c}
      P_0 & P_1 & P_2 \\
      P_2 & P_0 & P_1 \\
      P_1 & P_2 & P_0 \\
    \end{array}
  \right)^{n-1}
  \left(
    \begin{array}{c}
      1 - 3 P_0 \\
      1 - 3 P_2 \\
      1 - 3 P_1 \\
    \end{array}
  \right)\,.
  \label{v0vn1SMAhc}
\end{eqnarray}
Hence the diffusion coefficient (\ref{dcoef}) is
\begin{equation}
  \fl
  D_\mathrm{1SMA} = D_\mathrm{NMA}\left[1 +
    \Big(1\enspace 0 \enspace 0\Big)
    \left(
      \begin{array}{c@{\enspace}c@{\enspace}c}
        1 - P_0 & - P_1 & - P_2 \\
        - P_2 & 1 - P_0 & - P_1 \\
        - P_1 & - P_2 & 1 - P_0 \\
      \end{array}
    \right)^{-1}
    \left(
      \begin{array}{c}
        1 - 3 P_0 \\
        1 - 3 P_2 \\
        1 - 3 P_1 \\
      \end{array}
    \right)\right].
  \label{dcoef1SMAhc}
\end{equation}

Again, in the case of an isotropic process for which $P_2 = P_1 \eqdef
\ps$ (`symmetric'), equation (\ref{idthetan}) holds, and
substituting $\ps = (1- P_0)/2$ gives the following expression for 
the diffusion coefficient:
\begin{equation}
  D_\mathrm{1SMA} = D_\mathrm{NMA}\frac{1 - P_0 + \ps} {1 + P_0 - \ps}
  = D_\mathrm{NMA}\frac{3(1 - P_0)} {1 + 3 P_0}\,.
    \label{D1SMAHc}
\end{equation}
For an asymmetric process for which $P_1 \ne P_2$, defining the symmetric and
antisymmetric parts
$\ps \defeq (P_1 + P_2)/2$ and $\pa \defeq (P_1 - P_2)/2$,
we instead obtain
\begin{eqnarray}
  D_\mathrm{1SMA} &= D_\mathrm{NMA}
  \frac{1 - 3 \pa^2 - (P_0 - \ps)^2}{3 \pa^2 + (1 + P_0 - \ps)^2}\,,
  \nonumber\\
  &= D_\mathrm{NMA}\frac{3(1 - P_0)(1 + 3 P_0) - 12 \pa^2} 
  {(1 + 3 P_0)^2 + 12 \pa^2}\,.
  \label{D1SMAasymHc}
\end{eqnarray}

\subsection{\label{sub.1smat}Two-dimensional triangular lattice} 

Persistent walks on a triangular lattice were made popular by Fink and Mao
\cite{Fink1999} in
connection to tie knots. Here,
each site has $z=6$ neighbours, so that $\R$ is an anti-clockwise 
rotation by angle $\pi/3$. Following our convention [see figure
\ref{fig.num} (d)], we denote by $P_0 \defeq P(\vv|\vv)$ the
probability that the particle moves forward, $P_1 \defeq P(\R \vv|\vv)$
that it moves in the forward left direction relative to its previous
direction, and similarly for $P_2$, $P_3$, $P_4$, and $P_5$. 
 
Proceeding along the lines of the previous subsection, we let
\begin{equation}
  \PM^n \equiv 
  \left(
    \begin{array}{c@{\enspace}c@{\enspace}c@{\enspace}c}
      \an{n}{1} & \an{n}{2} & \cdots & \an{n}{6} \\
      \an{n}{6} & \an{n}{1} & \cdots & \an{n}{5}\\
      \vdots & \vdots & \ddots & \vdots\\
      \an{n}{2} & \an{n}{3} & \cdots & \an{n}{1}
    \end{array}
  \right)\,,
  \label{mnatr}
\end{equation}
in terms of which the velocity auto-correlation (\ref{v0vn1sma}) is given
by 
\begin{equation}
  \langle \vv_0\cdot\vv_n \rangle =
  \an{n}{1} + \frac{1}{2} [\an{n}{2} + \an{n}{6}
  - \an{n}{3} - \an{n}{5}] 
  - \an{n}{4} 
  \,.
\end{equation}
A computation similar to that of equation (\ref{eqmatrix}) yields
\begin{equation}
  \fl
  \langle \vv_0\cdot\vv_n \rangle = \Big(1\enspace 0   \enspace \cdots
  \enspace 0 \Big)
   \left(
    \begin{array}{c@{\enspace}c@{\enspace}c@{\enspace}c}
      P_0 & P_1 & \cdots & P_5 \\
      P_5 & P_0 & \cdots & P_4 \\
      \vdots & \vdots &\ddots & \vdots\\
      P_1 & P_2 & \cdots & P_0
    \end{array}
  \right)^{n-1}
  \left(
    \begin{array}{c}
      \ds P_0 - P_3 + \onehalf (P_1 - P_2 - P_4 + P_5)  \\
      \ds P_1 - P_4 + \onehalf (P_2 - P_3 - P_5 + P_0)  \\
      \vdots \\
      \ds P_5 - P_2 + \onehalf (P_0 - P_1 - P_3 + P_4)
    \end{array}
  \right)\,.
  \label{v0vn1SMAtr}
\end{equation}
The diffusion coefficient (\ref{dcoef}) is then
\begin{eqnarray}
  D_\mathrm{1SMA} =& D_\mathrm{NMA}\left[1 +
    \big(1\enspace 0 \enspace \cdots  \enspace 0\big)
    \left(
      \begin{array}{c@{\enspace}c@{\enspace}c@{\enspace}c}
        1 - P_0 & - P_1 & \cdots & - P_5 \\
        - P_5 & 1- P_0 & \cdots & - P_4 \\
        \vdots & \vdots & \ddots & \vdots\\
        - P_1 & - P_2 & \cdots & 1 - P_0
    \end{array}
    \right)^{-1} \right.
  \nonumber\\
  &\times
  \left.
    \left(
      \begin{array}{c}
        \ds P_0 - P_3 + \onehalf (P_1 - P_2 - P_4 + P_5)  \\
        \ds P_1 - P_4 + \onehalf (P_2 - P_3 - P_5 + P_0)  \\
        \vdots \\
        \ds P_5 - P_2 + \onehalf (P_0 - P_1 - P_3 + P_4)
      \end{array}
    \right)\right].
  \label{dcoef1SMAtr}
\end{eqnarray}

\subsection{\label{sub.1smadd}$d$-dimensional hypercubic lattice}

The case of a hypercubic lattice in arbitrary dimension $d$ with
coordination number $z = 2d$ can also be
treated, provided that the same probability
$\ps$ is assigned to scattering along all new directions which are
perpendicular to the previous direction of motion. We then have
$P_0 + P_{z/2} + 2(d-1) \ps = 1$, and the 
invariant distribution of velocities $p(\vv) = 1/(2d)$ is uniform. We
then recover the expression (\ref{D1SMASq}) for the diffusion coefficient in
this case.

\section{\label{sec.2sma}Two-Step Memory Approximation (2-SMA)}

We now turn to the main contribution of this paper, namely the calculation of
the diffusion coefficient for persistent random walks with $2$-step memory, for
several two-dimensional lattices.  

We thus assume that the velocity vectors obey a random process for which
the probability of $\vv_{n}$ takes on 
different values according to the velocities at the two previous steps,
$\vv_{n-1}$ and $\vv_{n-2}$, so that 
we may write
\begin{equation}
  \mu(\{\vv_{0},\ldots, \vv_{n}\}) = \prod_{i = 2}^n 
  P(\vv_{i}|\vv_{i-1}, \vv_{i-2}) p(\vv_{0},
  \vv_{1})\,.
\end{equation}
The velocity auto-correlation (\ref{v0vn}) function is then
\begin{equation}
  \langle   \vv_{0} \cdot \vv_{n} \rangle
  = \sum_{\{\vv_{n},\ldots,\vv_{0}\}}
  \vv_{0} \cdot \vv_{n} 
 \prod_{i = 2}^n 
  P(\vv_{i}|\vv_{i-1}, \vv_{i-2}) p(\vv_{0},
  \vv_{1})\,.
  \label{2spa0}
\end{equation}
The calculation of these correlations proceeds in the most
straightforward way by transposing the calculation leading to equation
(\ref{v0vn}) to the two-step probability transitions as characterizing the
probability transitions of a two-dimensional Markov chain. Considering a
lattice with coordination number $z$, the state of the Markov chain is a
normalised vector of dimension $z^2$.
The time evolution is specified by the $(z^2\times z^2)$
stochastic matrix $\PM$ with entries
\begin{equation}
  \Pm_{i = (i_1-1)z + i_2, j = (j_1-1)z + j_2}
  \defeq \delta_{i_2, j_{1}}
  P(\R^{i_1}\T^{2}\vv|\R^{j_1}\T\vv, 
  \R^{j_2}\vv)\,,
  \label{2dmarkov}
\end{equation}
where $i_1$, $i_2$,  $i_1$, and $i_2$ take values between $1$ and $z$.
Denoting by $\PP$ the invariant distribution of this Markov chain, i.e.\
the $z^2$-dimensional vector with components $\Pp_i$ such that $\sum_{j =
  1}^{z^2}\Pm_{i,j}\Pp_j = \Pp_i$, equation (\ref{2spa0}) becomes
\begin{equation}
  \langle   \vv_{0} \cdot \vv_{n} \rangle
  =  \sum_{i_0,i_n = 1}^{z^2}
  \D^{(n)}_{i_n,i_0}\,
  \mm{n-1}{i_n, i_0} \Pp_{i_0}\,,
  \label{v0vn2SMA}
\end{equation}
where $\D^{(n)}$ is the $(z^2\times z^2)$ matrix with elements 
\begin{equation}
  \D^{(n)}_{i =(i_1-1)z + i_2, j = (j_1-1)z + j_2} 
  \equiv \ee_{i_1} \cdot \T^n \ee_{j_2}\,,
  \label{Dmat}
\end{equation}
and $\ee_{k}$ denotes the unit vector along the $k$th lattice
direction.  

Using the symmetries of the problem and writing $P_{j\,k} = 
P(\R^{k}\T\R^{j}\T\vv|\R^{j}\T\vv, \vv)$ for the conditional probability of
turning successively by angles $2 \pi j/z$ and $2\pi k/z$ with respect to
the current direction (with reflection by $\T$ where needed), we define $\phi
\equiv \exp(2\ii\pi/z)$, where $\ii$ denotes the imaginary unit,  $\ii =
\sqrt{-1}$, and show, through the examples below, that equation
(\ref{v0vn2SMA}) reduces to the general expression  
\begin{eqnarray}
  \fl \langle   \vv_{0} \cdot \vv_{n} \rangle
  = \sigma^n \frac{z}{2}\Big( 1 \enspace \cdots \enspace 1\Big)
  \left[
    \left(
      \begin{array}{c@{\enspace}c@{\enspace}c@{\enspace}c}
        \ds P_{00} & \ds P_{10} & \ds \cdots & \ds P_{z-1,0}  \\
        \ds \phi P_{01} & \ds \phi P_{11} & \ds \cdots & \ds \phi
        P_{z-1,1} \\   
        \ds \vdots & \ds \vdots & \ds \ddots & \ds \vdots \\
        \ds \phi^{z-1} P_{0,z-1} & \ds \phi^{z-1} P_{1,z-1} & \ds \cdots
        & \ds  \phi^{z-1} P_{z-1,1}
      \end{array}
    \right)^{n-1}
    \hspace{-.5cm}
    \mathrm{Diag}(1, \phi, \dots, \phi^{z-1})
  \right.
  \nonumber\\
+
  \left. 
    \left(
      \begin{array}{c@{\enspace}c@{\enspace}c@{\enspace}c}
        \ds P_{00} & \ds P_{10} & \ds \cdots & \ds P_{z-1,0}  \\ 
        \ds \phi^{-1} P_{01} & \ds \phi^{-1} P_{11} & \ds \cdots & \ds
        \phi^{-1} 
        P_{z-1,1} \\   
        \ds \vdots & \ds \vdots & \ds \ddots & \ds \vdots \\
        \ds \phi^{1-z} P_{0,z-1} & \ds \phi^{1-z} P_{1,z-1} & \ds \cdots
        & \ds  \phi^{1-z} P_{z-1,1}
      \end{array}
    \right)^{n-1}
    \hspace{-.5cm}
    \mathrm{Diag}(1, \phi^{-1}, \dots, \phi^{1-z})
  \right]
  \left(
    \begin{array}{c}
      \Pp_1 \\
      \vdots \\
      \Pp_z
    \end{array}
  \right),\nonumber\\
  \label{v0vn2SMAgen}
\end{eqnarray}
where $\mathrm{Diag}(1, \phi, \dots, \phi^{z-1})$ denotes the
matrix with elements listed on the main diagonal and $0$ elsewhere. 
Here, $\sigma$ is a sign factor which is $-1$ for the honeycomb lattice and $+1$
for the other lattices, and reflects the action of $\T$.
Note that the second term in the summation is the complex conjugate of the
first, so that the result is real.
The diffusion coefficient for persistent random walks with $2$-step memory
is therefore 
\begin{eqnarray}
  \fl
  \frac{D_\mathrm{2SMA}}{D_\mathrm{NMA}}
  = 1 + \sigma
  z\Big( 1 \enspace \cdots \enspace 1\Big)\times
  \label{dcoef2SMAgen}\\  
\hspace{-1cm}
  \left[
    \left(
      \begin{array}{c@{\enspace}c@{\enspace}c@{\enspace}c}
        \ds 1 - P_{00} & \ds - P_{10} & \ds \cdots & \ds - 
        P_{z-1,0}  \\
        \ds - \phi P_{01} & \ds 1 - \phi P_{11} & \ds \cdots & 
        \ds - \phi P_{z-1,1} \\   
        \ds \vdots & \ds \vdots & \ds \ddots & \ds \vdots \\
        \ds - \phi^{z-1} P_{0,z-1} & \ds - \phi^{z-1} P_{1,z-1} & 
        \ds \cdots & \ds 1 - \phi^{z-1} P_{z-1,1}
      \end{array}
    \right)^{-1}
    \hspace{-.5cm}
    \mathrm{Diag}(1, \phi, \dots, \phi^{z-1})
  \right.
  \nonumber\\
  \hspace{-1cm}
  \left. +
    \left(
      \begin{array}{c@{\enspace}c@{\enspace}c@{\enspace}c}
        \ds 1 - P_{00} & \ds - P_{10} & \ds \cdots & 
        \ds - P_{z-1,0}  \\ 
        \ds - \phi^{-1} P_{01} & \ds - \phi^{-1} P_{11} & \ds \cdots 
        & \ds - \phi^{-1} P_{z-1,1} \\   
        \ds \vdots & \ds \vdots & \ds \ddots & \ds \vdots \\
        \ds - \phi^{1-z} P_{0,z-1} & \ds - \phi^{1-z} P_{1,z-1} 
        & \ds \cdots & \ds  1 - \phi^{1-z} P_{z-1,1}
      \end{array}
    \right)^{-1}
    \hspace{-.5cm}
    \mathrm{Diag}(1, \phi^{-1}, \dots, \phi^{1-z})
  \right]
  \left(
    \begin{array}{c}
      \Pp_1 \\
      \vdots \\
      \Pp_z
    \end{array}
  \right).\nonumber
\end{eqnarray} 

\subsection{\label{sub.2sma1d}One-dimensional lattice}

We first consider the simplest case, namely the one-dimensional lattice. 
The stochastic matrix $\PM$ from equation (\ref{2dmarkov}) is
the $4\times 4$ matrix
\begin{eqnarray}
  \PM &= 
  \left(
    \begin{array}{c@{\enspace}c@{\enspace}c@{\enspace}c}
      P(v|v,v) & P(v|v,-v)&0&0\\
      0&0&P(-v|v,-v)&P(-v|v,v)\\
      P(-v|v,v) & P(-v|v,-v)&0&0\\
      0&0&P(v|v,-v)&P(v|v,v)
    \end{array}
  \right),\nonumber\\
  &= \left(
    \begin{array}{c@{\enspace}c@{\enspace}c@{\enspace}c}
      P_{0\, 0} & P_{1\, 0} & 0 & 0\\
      0 & 0 & P_{1\, 1} & P_{0\, 1}\\
      P_{0\, 1} & P_{1\, 1} & 0 & 0\\
      0 & 0 & P_{1\, 0} & P_{0\, 0}
    \end{array}
  \right)\,,
\end{eqnarray}
where $P_{0\, 0} + P_{0\, 1} = 1$ and  $P_{1\, 0} + P_{1\, 1} = 1$.

Considering equation (\ref{v0vn2SMA}), we compute the invariant
distribution of $\PM$, which is the vector $\PP$ whose components
correspond to the four states $p(v,v),\, p(-v,v),\, p(v,-v)$ and
$p(-v,-v)$. Given that we must have  $p(-v,-v) = p(v,v)$ and $p(v,-v) =
p(-v,v)$, the equilibrium distribution is obtained as the solution of the
system of equations 
\begin{equation}
  \eqalign{
    p(v,v) = P_{00}\, p(v,v) + P_{10}\, p(-v,v)\,,\\
    p(v,v) + p(-v,v) = \frac{1}{2}\,,
  }
\end{equation}
giving
\begin{equation}
  \eqalign{
    \Pp_1 = \Pp_4 = p(v,v) = 
    \frac{P_{10}}{2[1 - P_{00} + P_{10}]}\,,\\
    \Pp_2 = \Pp_3 = p(-v,v) = 
    \frac{1 - P_{00}}{2[1 - P_{00} + P_{10}]}\,.\\
  }
\end{equation}
The matrix $\D^{(n)}$, equation (\ref{Dmat}), is here the same for all
$n$, and has the expression
\begin{equation}
  \D = 
  \left(
    \begin{array}{c@{\enspace}c@{\enspace}c@{\enspace}c}
      1 & -1&1&-1\\
      1 & -1&1&-1\\
      -1 & 1& -1&1\\
      -1 & 1& -1&1
    \end{array}
  \right)\,.
\end{equation}

The velocity auto-correlation (\ref{v0vn2SMA}) is thus
\begin{eqnarray}
  \fl 
  \langle \vv_0 \cdot \vv_{n} \rangle =
  2 \Big[\left(\mn{n-1}{1,1} - \mn{n-1}{1,4}  + \mn{n-1}{3,4}
    - \mn{n-1}{3,1}\right)
  \Pp_1 
  \nonumber\\
  + \left(\mn{n-1}{1,3} - \mn{n-1}{1,2} + \mn{n-1}{3,2}
    - \mn{n-1}{3,3} \right)
    \Pp_2\Big]\,, 
  \nonumber\\
  \lo=
  2\, \Big(1\enspace 1\Big) 
  \left(
    \begin{array}{c@{\enspace}c}
      \mn{n-1}{1,1} - \mn{n-1}{1,4} & \mn{n-1}{1,3} - \mn{n-1}{1,2}\\
      \mn{n-1}{3,4} - \mn{n-1}{3,1} & \mn{n-1}{3,2} - \mn{n-1}{3,3}
    \end{array}
  \right) 
  \left(
    \begin{array}{c}
      \Pp_1\\
      \Pp_2
    \end{array}
  \right)\,.
  \label{v0vn2SMAexp}
\end{eqnarray}

Since $\PM^n$, the $n$th power of $\PM$, has the symmetries of $\PM$,
its entries $\mn{n}{i,j}$ are such that
\begin{equation}
  \eqalign{
    \mn{n}{1,1} = \mn{n}{4,4} \eqdef \an{n}{1}\,,\qquad
    \mn{n}{2,1} = \mn{n}{3,4} \eqdef \bn{n}{4}\,,\\
    \mn{n}{1,2} = \mn{n}{4,3} \eqdef \an{n}{2}\,,\qquad
    \mn{n}{2,2} = \mn{n}{3,3} \eqdef \bn{n}{3} \,,\\
    \mn{n}{1,3} = \mn{n}{4,2} \eqdef \an{n}{3}\,,\qquad
    \mn{n}{2,3} = \mn{n}{3,2} \eqdef \bn{n}{2}\,,\\
    \mn{n}{1,4} = \mn{n}{4,1} \eqdef \an{n}{4}\,,\qquad
    \mn{n}{2,4} = \mn{n}{3,1} \eqdef \bn{n}{1}\,.
  }
  \label{mnab}
\end{equation}
Writing $\PM^n = \PM \PM^{n-1}$, we obtain two separate sets of equations
for $2\times 2$ matrices, one involving $\an{n}{1}, \an{n}{4}$ and
$\bn{n}{1}, \bn{n}{4}$, and the other  involving $\an{n}{2}, \an{n}{3}$ and
$\bn{n}{2}, \bn{n}{3}$: 
\begin{equation}
  \eqalign{
    \left(
      \begin{array}{c@{\enspace}c}
        \an{n}{1} & \an{n}{4}\\
        \bn{n}{1} & \bn{n}{4}
      \end{array}
    \right) = 
    \left(
      \begin{array}{c@{\enspace}c}
        P_{0\, 0} & P_{1\, 0}\\
        P_{0\, 1} & P_{1\, 1}
      \end{array}
    \right)
    \left(
      \begin{array}{c@{\enspace}c}
        \an{n-1}{1} & \an{n-1}{4}\\
        \bn{n-1}{4} & \bn{n-1}{1}
      \end{array}
    \right)\,,\\
    \left(
      \begin{array}{c@{\enspace}c}
        \an{n}{2} & \an{n}{3}\\
        \bn{n}{2} & \bn{n}{3}
      \end{array}
    \right) = 
    \left(
      \begin{array}{c@{\enspace}c}
        P_{0\, 0} & P_{1\, 0}\\
        P_{0\, 1} & P_{1\, 1}
      \end{array}
    \right)
    \left(
      \begin{array}{c@{\enspace}c}
        \an{n-1}{2} & \an{n-1}{3}\\
        \bn{n-1}{3} & \bn{n-1}{2}
      \end{array}
    \right)\,.
  }
\end{equation}
Note that these equations do not have a simple recursive form, since the
elements of the matrices on the two sides do not appear in the same
places. However, taking the differences $\an{n}{1} - \an{n}{4}$, $\an{n}{3}
- \an{n}{2}$,  $\bn{n}{4} - \bn{n}{1}$, and  $\bn{n}{2} - \bn{n}{3}$,
we obtain the recursive system
\begin{eqnarray}
  \fl
  \left(
    \begin{array}{c@{\enspace}c}
      \an{n}{1} - \an{n}{4} & \an{n}{3} - \an{n}{2}\\
      \bn{n}{4} - \bn{n}{1} & \bn{n}{2} - \bn{n}{3}
    \end{array}
  \right) \nonumber\\
  = \left(
    \begin{array}{c@{\enspace}c}
      P_{0\, 0} & P_{1\, 0}\\
      - P_{0\, 1} & - P_{1\, 1}
    \end{array}
  \right)
  \left(
    \begin{array}{c@{\enspace}c}
      \an{n-1}{1} - \an{n-1}{4} & \an{n-1}{3} - \an{n-1}{2}\\
      \bn{n-1}{4} - \bn{n-1}{1} & \bn{n-1}{2} - \bn{n-1}{3}
    \end{array}
  \right)\,,\nonumber\\
  = \left(
    \begin{array}{c@{\enspace}c}
      P_{0\, 0} &  P_{1\, 0}\\
      - P_{0\, 1} & - P_{1\, 1}
    \end{array}
  \right)^{n-1}
  \left(
    \begin{array}{c@{\enspace}c}
      \an{1}{1} - \an{1}{4} & \an{1}{3} - \an{1}{2}\\
      \bn{1}{4} - \bn{1}{1} & \bn{1}{2} - \bn{1}{3}
    \end{array}
  \right)\,,\nonumber\\
  = \left(
    \begin{array}{c@{\enspace}c}
      P_{0\, 0} &  P_{1\, 0}\\
      - P_{0\, 1} & - P_{1\, 1}
    \end{array}
  \right)^{n-1}
  \left(
    \begin{array}{c@{\enspace}c}
      P_{0\, 0} &  -P_{1\, 0}\\
      - P_{0\, 1} & P_{1\, 1}
    \end{array}
  \right)\,,\nonumber\\
  = \left(
    \begin{array}{c@{\enspace}c}
      P_{0\, 0} &  P_{1\, 0}\\
      - P_{0\, 1} & - P_{1\, 1}
    \end{array}
  \right)^n
  \left(
    \begin{array}{c@{\enspace}c}
      1 & 0\\
      0 & -1
    \end{array}
  \right)\,.
  \label{recursionab}
\end{eqnarray}

Plugging this equation into equation (\ref{v0vn2SMAexp}), we obtain
\begin{equation}
  \langle \vv_0 \cdot \vv_{n} \rangle =
  2\, \Big(1 \enspace 1\Big) 
  \left(
    \begin{array}{c@{\enspace}c}
      P_{0\, 0} &  P_{1\, 0}\\
      - P_{0\, 1} & - P_{1\, 1}
    \end{array}
  \right)^{n-1}
  \left(
    \begin{array}{c}
      \Pp_1\\
      -\Pp_2
    \end{array}
  \right)\,.
  \label{v0vn2SMA1d}
\end{equation}
This is equation (\ref{v0vn2SMAgen}), where $\phi =
\exp(2\ii\pi/2) = -1$. The diffusion coefficient is therefore given by
equation (\ref{dcoef2SMAgen}):
\begin{eqnarray}
  \frac{D_\mathrm{2SMA}}{D_\mathrm{NMA}}
  &=&  
  1 + 4 \Big(1 \enspace 1\Big) 
  \left(
    \begin{array}{c@{\enspace}c}
      1 - P_{00} & - P_{10}\\
      P_{01} & 1 + P_{11}
    \end{array}
  \right)^{-1}
  \left(
    \begin{array}{c}
      \Pp_1\\
      -\Pp_2
    \end{array}
  \right)\,,\nonumber\\
  &=& D_\mathrm{NMA}\frac{P_{10}}{[1 - P_{00}]}
  \frac{[1 + P_{00} - P_{10}]}
  {[1 - P_{00} + P_{10}]}\,.
  \label{dcoef2spa1d}
\end{eqnarray}
It is a function
of the two parameters $P_{00}$ and $P_{10}$; when these are equal, the
process reduces to a walk with single-step memory, 
and the diffusion coefficient to that of the single-step memory
approximation (\ref{dcoef1spa1d}), as it should. 

\subsection{\label{sub.2smahc}Two-dimensional honeycomb lattice}

For the two-dimensional honeycomb lattice, 
the stochastic matrix $\PM$ of equation (\ref{2dmarkov}) is a
$(9\times 9)$ matrix with the following non-zero entries:
\begin{equation}
  \eqalign{
    \Pm_{1,1} = \Pm_{5,5} = \Pm_{9,9} = P(\vv| -\vv, \vv)
    = P_{00}\,,\\
    \Pm_{1,2} = \Pm_{5,6} = \Pm_{9,7} = P(\R^2\vv| -\R^2\vv,\vv) 
    = P_{10}\,,\\
    \Pm_{1,3} = \Pm_{5,4} = \Pm_{9,8} = P(\R\vv| -\R\vv,\vv) 
    = P_{20}\,,\\ 
    \Pm_{4,1} = \Pm_{8,5} = \Pm_{3,9} = P(\R\vv| -\vv, \vv)
    = P_{02}\,,\\
    \Pm_{4,2} = \Pm_{8,6} = \Pm_{3,7} = P(\vv| -\R^2\vv, \vv)
    = P_{12}\,,\\
    \Pm_{4,3} = \Pm_{8,4} = \Pm_{3,8} = P(\R^2\vv| -\R\vv, \vv)
    = P_{22}\,,\\
    \Pm_{7,1} = \Pm_{2,5} = \Pm_{6,9} = P(\R^2\vv| -\vv, \vv)
    = P_{01}\,,\\
    \Pm_{7,2} = \Pm_{2,6} = \Pm_{6,7} = P(\R\vv| -\R^2\vv, \vv)
    = P_{11}\,,\\
    \Pm_{7,3} = \Pm_{2,4} = \Pm_{6,8} = P(\vv| -\R\vv, \vv)
    = P_{21}\,.
  }
\end{equation}
Given the three constraints 
\begin{equation}
  \eqalign{
    P_{00} + P_{01} + P_{02} = 1\,,\\
    P_{10} + P_{11} + P_{12} = 1\,,\\ 
    P_{20} + P_{21} + P_{22} = 1\,,
  }
\end{equation}
the actual number of independent parameters is six. 

The invariant distribution $\PP$ with components $\Pp_i$ can be written in
terms of the three probabilities $p(\vv, -\vv)$, $p(\vv, -\R\vv)$, $p(\vv,
-\R^2\vv)$,  
\begin{equation}
  \eqalign{
    \Pp_1 = \Pp_5 = \Pp_9 = p(\vv, -\vv)\,,\\
    \Pp_2 = \Pp_6 = \Pp_7 = p(\vv, -\R\vv)\,,\\
    \Pp_3 = \Pp_4 = \Pp_8 = p(\vv, -\R^2\vv)\,,
  }
\end{equation}
which are solutions of the system of equations:
\begin{eqnarray}
  p(\vv, -\vv) =  P_{00}\, p(\vv, -\vv) + P_{10}\, p(\vv, -\R^2\vv) 
  + P_{20}\, p(\vv, -\R\vv)\,,\\
  p(\vv, -\R^2\vv) =  P_{01}\, p(\vv, -\vv) + P_{11}\, p(\vv, -\R^2\vv) 
  +  P_{21}\, p(\vv, -\R\vv)\,,\\
  p(\vv,-\vv) + p(\vv, -\R\vv) + p(\vv, -\R^2\vv)
  = \frac{1}{3}\,.
\end{eqnarray}
The matrix (\ref{Dmat}) has the block structure
\begin{equation}
  \D^{(n)} = (-1)^n
  \left(\begin{array}{c@{\enspace}c@{\enspace}c}
      \B_{1} & \B_{1} &\B_{1}\\
      \B_{2} & \B_{2} &\B_{2}\\
      \B_{3} & \B_{3} &\B_{3}
    \end{array}
  \right)\,,
\end{equation}
where 
\begin{equation}
  \fl
  \B_1 = 
  \left(\begin{array}{c@{\enspace}c@{\enspace}c}
      -1&1/2&1/2\\
      -1&1/2&1/2\\
      -1&1/2&1/2
    \end{array}
  \right),
  \B_2 = 
  \left(\begin{array}{c@{\enspace}c@{\enspace}c}
      1/2&-1&1/2\\
      1/2&-1&1/2\\
      1/2&-1&1/2
    \end{array}
  \right),
  \B_3 = 
  \left(\begin{array}{c@{\enspace}c@{\enspace}c}
      1/2&1/2&-1\\
      1/2&1/2&-1\\
      1/2&1/2&-1
    \end{array}
  \right).
\end{equation}

Substituting these expressions into equation (\ref{v0vn2SMA}), we find
\begin{eqnarray}
  \fl\langle \vv_{n+1}\cdot\vv_0\rangle
  = 3 (-1)^{n+1}
  \Bigg\{
  \Big[\mn{n}{1,1} - \frac{1}{2} \mn{n}{1,5} - \frac{1}{2} \mn{n}{1,9} 
  - \frac{1}{2} \mn{n}{4,1} + \mn{n}{4,5} - \frac{1}{2} \mn{n}{4,9} 
  - \frac{1}{2} \mn{n}{7,1} 
  \nonumber\\
  - \frac{1}{2} \mn{n}{7,5} + \mn{n}{7,9}\Big] \Pp_1 
  + \Big[- \frac{1}{2} \mn{n}{1,2} - \frac{1}{2} \mn{n}{1,6} + \mn{n}{1,7} 
  + \mn{m,n}{4,2} - \frac{1}{2} \mn{n}{4,6} 
  \nonumber \\
- \frac{1}{2} \mn{n}{4,7} 
  - \frac{1}{2} \mn{n}{7,2} + \mn{n}{7,6} - \frac{1}{2} \mn{n}{7,7}\Big] 
  \Pp_2 
  + \Big[- \frac{1}{2} \mn{n}{1,3} + \mn{n}{1,4} - \frac{1}{2} \mn{n}{1,8} 
  \nonumber \\
  - \frac{1}{2} \mn{n}{4,3} - \frac{1}{2} \mn{n}{4,4} 
  + \mn{n}{4,8} 
  + \mn{n}{7,3} - \frac{1}{2} \mn{n}{7,4} - \frac{1}{2}
  \mn{n}{7,8}\Big] \Pp_3\Bigg\} \,,\nonumber\\
  \lo = 3 (-1)^{n+1} \Big(1\enspace 1\enspace 1\Big)  \times
  \label{v0vn2SMAhc1}\\
  \fl 
  \left(
        \begin{array}{c@{\enspace}c@{\enspace}c}
      \mn{n}{1,1} - \frac{1}{2} \mn{n}{1,5} - \frac{1}{2} \mn{n}{1,9}&
      - \frac{1}{2} \mn{n}{1,2} - \frac{1}{2} \mn{n}{1,6} + \mn{n}{1,7}&
      - \frac{1}{2} \mn{n}{1,3} + \mn{n}{1,4} - \frac{1}{2} \mn{n}{1,8}\\
      - \frac{1}{2} \mn{n}{7,1} - \frac{1}{2} \mn{n}{7,5} + \mn{n}{7,9}&
      - \frac{1}{2} \mn{n}{7,2} + \mn{n}{7,6} - \frac{1}{2} \mn{n}{7,7}&
      \mn{n}{7,3} - \frac{1}{2} \mn{n}{7,4} - \frac{1}{2} \mn{n}{7,8}\\
      - \frac{1}{2} \mn{n}{4,1} + \mn{n}{4,5} - \frac{1}{2} \mn{n}{4,9}&
      \mn{n}{4,2} - \frac{1}{2} \mn{n}{4,6} - \frac{1}{2} \mn{n}{4,7}&
      - \frac{1}{2} \mn{n}{4,3} - \frac{1}{2} \mn{n}{4,4} + \mn{n}{4,8}
    \end{array}
  \right)
  \left(
    \begin{array}{c}
      \Pp_1\\
      \Pp_2\\
      \Pp_3
    \end{array}
  \right)\nonumber\,.
\end{eqnarray}

Proceeding along the lines of the computation presented in subsection
\ref{sub.2sma1d}, we obtain a set of recursive matrix equations
(\ref{recursionabcapp}) involving the coefficients of $\PM^n$. We refer the
reader to \ref{app.hc} for the details of this derivation. 

We note that the coefficients which appear in equation
(\ref{v0vn2SMAhc1}) satisfy identities such as, for instance, 
\begin{eqnarray}
  \mn{n}{1,1} - \frac{1}{2} \mn{n}{1,5} - \frac{1}{2} \mn{n}{1,9} 
  =& \frac{1}{2}\left[\mn{n}{1,1} + e^{2\ii\pi/3} \mn{n}{1,5} +
    e^{-2\ii\pi/3} \mn{n}{1,9}\right] 
  \nonumber\\
  &+\frac{1}{2}\left[\mn{n}{1,1} + e^{-2\ii\pi/3} \mn{n}{1,5} +
    e^{2\ii\pi/3} \mn{n}{1,9}\right]\,.
\end{eqnarray}
Thus, letting $\phi = \exp(2\ii\pi/3)$, 
we can combine the results of equation (\ref{recursionabcapp})
with equation (\ref{v0vn2SMAhc1}) to find:
\begin{eqnarray}
  \fl\langle \vv_{n}\cdot\vv_0\rangle
  =  (-1)^{n} \frac{3}{2}
  \Big(1\enspace 1\enspace 1\Big)
  \left[
  \left(
    \begin{array}{c@{\enspace}c@{\enspace}c}
      P_{00}& P_{10}& P_{20} \\
      \phi P_{01}& \phi P_{11}& \phi P_{21} \\
      \phi^2 P_{02}& \phi^2 P_{12}& \phi^2 P_{22} 
    \end{array}
  \right)^{n-1}
  \left(
    \begin{array}{c@{\enspace}c@{\enspace}c}
      1 & 0 & 0 \\
      0 & \phi & 0 \\
      0 & 0 & \phi^2
    \end{array}
  \right)
  \right.
  \nonumber\\
  +\left.
  \left(
    \begin{array}{c@{\enspace}c@{\enspace}c}
      P_{00}& P_{10}& P_{20} \\
      \phi^2 P_{01}& \phi^2 P_{11}& \phi^2 P_{21} \\
      \phi P_{02}& \phi P_{12}& \phi P_{22} 
    \end{array}
  \right)^{n-1}
  \left(
    \begin{array}{c@{\enspace}c@{\enspace}c}
      1 & 0 & 0 \\
      0 & \phi^2 & 0 \\
      0 & 0 & \phi
    \end{array}
  \right)\right]
  \left(
    \begin{array}{c}
      \Pp_1\\
      \Pp_2\\
      \Pp_3
    \end{array}
  \right)\, .
\end{eqnarray}
This is equation (\ref{v0vn2SMAgen}).
The
diffusion coefficient (\ref{dcoef}) is therefore given by
(\ref{dcoef2SMAgen}), which is here  
\begin{eqnarray}
  \fl
  \frac{D_\mathrm{2SMA}}{D_\mathrm{NMA}}
  = 
  1 - 3
  \Big(1\enspace 1\enspace 1\Big)
  \left[
  \left(
    \begin{array}{c@{\enspace}c@{\enspace}c}
      1 + P_{00}& P_{10}& P_{20} \\
      \phi P_{01}& 1 + \phi P_{11}& \phi P_{21} \\
      \phi^2 P_{02}& \phi^2 P_{12}& 1 + \phi^2 P_{22} 
    \end{array}
  \right)^{-1}
  \left(
    \begin{array}{c@{\enspace}c@{\enspace}c}
      1 & 0 & 0 \\
      0 & \phi & 0 \\
      0 & 0 & \phi^2
    \end{array}
  \right)
  \right.
  \nonumber\\
  \lo +\left.
  \left(
    \begin{array}{c@{\enspace}c@{\enspace}c}
      1 + P_{00}& P_{10}& P_{20} \\
      \phi^2 P_{01}& 1 + \phi^2 P_{11}& \phi^2 P_{21} \\
      \phi P_{02}& \phi P_{12}& 1 + \phi P_{22} 
    \end{array}
  \right)^{-1}
  \left(
    \begin{array}{c@{\enspace}c@{\enspace}c}
      1 & 0 & 0 \\
      0 & \phi^2 & 0 \\
      0 & 0 & \phi
    \end{array}
  \right)\right]
  \left(
    \begin{array}{c}
      \Pp_1\\
      \Pp_2\\
      \Pp_3
    \end{array}
  \right).
  \label{dcoef2spahc}
\end{eqnarray}

Given a symmetric process for which left and right probabilities are equal,
but the probability of a left--left turn is different than that of a
right--left turn, we let
\begin{equation}
  \eqalign{
    P_{02} = P_{01} \defeq \pbs 
    = \frac{1 - P_{00}} {2}\,,\\
    P_{10} = P_{20} \defeq \psb\,, 
    P_{11} = P_{22} \defeq \pss\,, 
    P_{12} = P_{21} = 1 - \psb - \pss\,.
  }
\end{equation}
Carrying out the matrix inversions in (\ref{dcoef2spahc}), we find the
diffusion coefficient,   
\begin{equation}
 \fl D_\mathrm{2SMA} = D_\mathrm{NMA}
 \frac{3 (1 - P_{00}) (1 + P_{00} - \psb) (2 - \psb - 2 \pss)}
  {(1 - P_{00} + \psb) 
    [\psb (7 + P_{00} - 8 \pss) + 2 (1 + P_{00}) \pss -4 \psb^2 ]}\,.
  \label{dcoef2spahcsym1}
\end{equation}
If we further assume complete left--right symmetry and identify the
probabilities of left--left turns and left--right turns, thus letting
\begin{equation}
  P_{12} = P_{21} = P_{11} = P_{22} = \frac{1 - \psb}{2}\,,
\end{equation}
equation (\ref{dcoef2spahcsym1}) simplifies to
\begin{equation}
  D_\mathrm{2SMA} = D_\mathrm{NMA}
  \frac{3 (1 - P_{00}) (1 + P_{00} - \psb)}
  {(1 - P_{00} + \psb) (1 + P_{00} + 2 \psb)}\,.
  \label{dcoef2spahcsym}
\end{equation}

\subsection{\label{sub.2smasq}Two-dimensional square lattice}

For the two-dimensional square lattice, 
the matrix $\PM$ of equation (\ref{2dmarkov}) is the $(16\times16)$
matrix with the following non-zero entries:
\begin{eqnarray*}
    \Pm_{1,1} &= \Pm_{6,6} = \Pm_{11,11} = \Pm_{16,16} 
    = P(\vv| \vv, \vv) 
    = P_{00}\,,\\
    \Pm_{1,2} &= \Pm_{6,7} = \Pm_{11,12} = \Pm_{16,13} 
    = P(\R^3\vv| \R^3\vv, \vv) = P_{10}\,,\\
    \Pm_{1,3} &= \Pm_{6,8} = \Pm_{11,9} = \Pm_{16,14} 
    = P(\R^2\vv| \R^2\vv, \vv) = P_{20}\,,\\
    \Pm_{1,4} &= \Pm_{6,5} = \Pm_{11,10} = \Pm_{16,15} 
    = P(\R\vv| \R\vv, \vv) = P_{30}\,,\\
    \Pm_{5,1} &= \Pm_{10,6} = \Pm_{15,11} = \Pm_{4,16} 
    = P(\R\vv| \vv, \vv) = P_{03}\,,\\
    \Pm_{5,2} &= \Pm_{10,7} = \Pm_{15,12} = \Pm_{4,13} 
    = P(\vv| \R^3\vv, \vv) = P_{13}\,,\\
    \Pm_{5,3} &= \Pm_{10,8} = \Pm_{15,9} = \Pm_{4,14} 
    = P(\R^3\vv| \R^2\vv, \vv) = P_{23}\,,\\
    \Pm_{5,4} &= \Pm_{10,5} = \Pm_{15,10} = \Pm_{4,15} 
    = P(\R^2\vv| \R\vv, \vv) = P_{33}\,,\\
    \Pm_{9,1} &= \Pm_{14,6} = \Pm_{3,11} = \Pm_{8,16} 
    = P(\R^2\vv| \vv, \vv) = P_{02}\,,\\
    \Pm_{9,2} &= \Pm_{14,7} = \Pm_{3,12} = \Pm_{8,13} 
    = P(\R\vv| \R^3\vv, \vv) = P_{12}\,,\\
    \Pm_{9,3} &= \Pm_{14,8} = \Pm_{3,9} = \Pm_{8,14} 
    = P(\vv| \R^2\vv, \vv) = P_{22}\,,\\
    \Pm_{9,4} &= \Pm_{14,5} = \Pm_{3,10} = \Pm_{8,15} 
    = P(\R^3\vv| \R\vv, \vv) = P_{32}\,,\\
    \Pm_{13,1} &= \Pm_{2,6} = \Pm_{7,11} = \Pm_{12,16} 
    = P(\R^3\vv| \vv, \vv) = P_{01}\,,\\
    \Pm_{13,2} &= \Pm_{2,7} = \Pm_{7,12} = \Pm_{12,13} 
    = P(\R^2\vv| \R^3\vv, \vv) = P_{11}\,,\\
    \Pm_{13,3} &= \Pm_{2,8} = \Pm_{7,9} = \Pm_{12,14} 
    = P(\R\vv| \R^2\vv, \vv) = P_{21}\,,\\
    \Pm_{13,4} &= \Pm_{2,5} = \Pm_{7,10} = \Pm_{12,15} 
    = P(\vv| \R\vv, \vv) = P_{31}\,.
\end{eqnarray*}

The matrix (\ref{Dmat}) is here
\begin{equation}
  \D^{(n)} \eqdef   \D = 
  \left(\begin{array}{c@{\enspace}c@{\enspace}c@{\enspace}c}
      \B_{1} & \B_{1} &\B_{1} & \B_{1}\\
      \B_{2} & \B_{2} &\B_{2} & \B_{2}\\
      - \B_{1} & - \B_{1} & - \B_{1} & - \B_{1}\\
      - \B_{2} & - \B_{2} & - \B_{2} & - \B_{2}
    \end{array}
  \right)\,,
\end{equation}
where 
\begin{equation}
  \B_1 = 
  \left(\begin{array}{c@{\enspace}c@{\enspace}c@{\enspace}c}
      1& 0 & -1 & 0\\
      1& 0 & -1 & 0\\
      1& 0 & -1 & 0\\
      1& 0 & -1 & 0
    \end{array}
  \right),\quad
  \B_2 = 
  \left(\begin{array}{c@{\enspace}c@{\enspace}c@{\enspace}c}
      0 & 1& 0 & -1\\
      0 & 1& 0 & -1\\
      0 & 1& 0 & -1\\
      0 & 1& 0 & -1
    \end{array}
  \right),
\end{equation}

The invariant distribution $\PP$ will not be written explicitly here. Due to
symmetry, only $4$ of the $16$ components are distinct. These four
components are most simply computed as the invariant vector of the following
$(4\times4)$ matrix:
\begin{equation}
  \left(
    \begin{array}{c@{\enspace}c@{\enspace}c@{\enspace}c}
      P_{00} & P_{10} & P_{20} & P_{30}\\
      P_{01} & P_{11} & P_{21} & P_{31}\\
      P_{02} & P_{12} & P_{22} & P_{32}\\
      P_{03} & P_{13} & P_{23} & P_{33}
    \end{array}
  \right)
  \left(
    \begin{array}{c}
      \Pp_1\\
      \Pp_2\\
      \Pp_3\\
      \Pp_4
    \end{array}
  \right)
  =   \left(
    \begin{array}{c}
      \Pp_1\\
      \Pp_2\\
      \Pp_3\\
      \Pp_4
    \end{array}
  \right)\,,
\end{equation}
normalised so that $\Pp_1 + \Pp_2 + \Pp_3 + \Pp_4 = 1/4$.

In terms of these quantities, the velocity auto-correlations are found to
be 
\begin{eqnarray}
  \fl
  \lo \langle \vv_0 \cdot \vv_{n+1} \rangle = 
  4 \left[\mn{n}{1,1} - \mn{n}{1,11} + \mn{n}{5,6} - \mn{n}{5,16} - 
    \mn{n}{9,1} + \mn{n}{9,11} - \mn{n}{13,6} + \mn{n}{13,16}\right] \Pp_1 
  \nonumber\\
  + 4 \left[-\mn{n}{1,7} + \mn{n}{1,13} + \mn{n}{5,2} - \mn{n}{5,12}   
    + \mn{n}{9,7}  - \mn{n}{9,13}  - \mn{n}{13,2} + \mn{n}{13,12}\right] \Pp_2 
  \nonumber\\
  + 4 \left[-\mn{n}{1,3} + \mn{n}{1,9} - \mn{n}{5,8} + \mn{n}{5,14} +
    \mn{n}{9,3} - \mn{n}{9,9} + \mn{n}{13,8} - \mn{n}{13,14}\right] \Pp_3 
  \nonumber\\
  + 4 \left[\mn{n}{1,5} - \mn{n}{1,15} - \mn{n}{5,4} + \mn{n}{5,10} - 
    \mn{n}{9,5} + \mn{n}{9,15} + \mn{n}{13,4} - \mn{n}{13,10}\right]
  \Pp_4\,,
  \nonumber\\
  \fl 
  = 4\Big(1\enspace 1\enspace 1\enspace 1\Big)
  \left(
    \begin{array}{c@{\enspace}c@{\enspace}c@{\enspace}c}
      \mn{n}{1,1} - \mn{n}{1,11} &
      \mn{n}{1,13} -\mn{n}{1,7} &
      \mn{n}{1,9} -\mn{n}{1,3} &
      \mn{n}{1,5} - \mn{n}{1,15} \\
      \mn{n}{13,16} - \mn{n}{13,6} &
      \mn{n}{13,12} - \mn{n}{13,2} &
      \mn{n}{13,8} - \mn{n}{13,14} &
      \mn{n}{13,4} - \mn{n}{13,10} \\
      \mn{n}{9,11} - \mn{n}{9,1} &
      \mn{n}{9,7}  - \mn{n}{9,13} &
      \mn{n}{9,3} - \mn{n}{9,9} & 
      \mn{n}{9,15} - \mn{n}{9,5} \\
      \mn{n}{5,6} - \mn{n}{5,16} &
      \mn{n}{5,2} - \mn{n}{5,12} &
      \mn{n}{5,14} - \mn{n}{5,8} &
      \mn{n}{5,10} - \mn{n}{5,4} 
    \end{array}
  \right)
  \left(
    \begin{array}{c}
      \Pp_1\\
      \Pp_2\\
      \Pp_3\\
      \Pp_4
    \end{array}
  \right)\,.
  \nonumber
\end{eqnarray}
Using the results of \ref{app.sq}, this expression reduces to
\begin{eqnarray}
  \fl \langle \vv_0 \cdot \vv_{n} \rangle = 
  2\Big(1\enspace 1\enspace 1\enspace 1\Big)\left[
    \left(
      \begin{array}{c@{\enspace}c@{\enspace}c@{\enspace}c}
        P_{00} & P_{10} & P_{20} & P_{30} \\
        \ii P_{01} & \ii P_{11} & \ii P_{21} & \ii P_{31} \\
        - P_{02} & - P_{12} & - P_{22} & - P_{32} \\
        -\ii P_{03} & -\ii P_{13} & -\ii P_{23} & -\ii P_{33} 
      \end{array}
    \right)^{n-1}
    \left(
      \begin{array}{c@{\enspace}c@{\enspace}c@{\enspace}c}
        1 & 0 & 0 & 0 \\
        0 & \ii & 0 & 0 \\
        0 & 0 & -1 & 0 \\
        0 & 0 & 0 & -\ii
      \end{array}
    \right)
    \right.
    \label{v0vn2smasqexp}\\
    +\left.
      \left(
        \begin{array}{c@{\enspace}c@{\enspace}c@{\enspace}c}
          P_{00} & P_{10} & P_{20} & P_{30} \\
          -\ii P_{01} & -\ii P_{11} & -\ii P_{21} & -\ii P_{31} \\
          -1 P_{02} & -1 P_{12} & -1 P_{22} & -1 P_{32} \\
          \ii P_{03} & \ii P_{13} & \ii P_{23} & \ii P_{33} 
        \end{array}
      \right)^{n-1}
      \left(
        \begin{array}{c@{\enspace}c@{\enspace}c@{\enspace}c}
          1 & 0 & 0 & 0 \\
          0 & -\ii & 0 & 0 \\
          0 & 0 & -1 & 0 \\
          0 & 0 & 0 & -\ii
        \end{array}
      \right)
    \right]
  \left(
    \begin{array}{c}
      \Pp_1\\
      \Pp_2\\
      \Pp_3\\
      \Pp_4
    \end{array}
  \right),
  \nonumber
\end{eqnarray}
which is equation (\ref{v0vn2SMAgen}). Substituting this result into
equation (\ref{dcoef}), we recover the diffuson coefficient given by
equation (\ref{dcoef2SMAgen})~: 
\begin{eqnarray}
  \fl\frac{D_\mathrm{2SMA}}{D_\mathrm{NMA}}
  = 1 + 4 \Big(1\enspace 1\enspace 1\enspace 1\Big)
  \times
  \label{dcoef2spa2dql}\\
  \left[
    \left(
      \begin{array}{c@{\enspace}c@{\enspace}c@{\enspace}c}
        1-P_{00} & -P_{10} & -P_{20} & -P_{30} \\
        -\ii P_{01} & 1-\ii P_{11} & -\ii P_{21} & -\ii P_{31} \\
        P_{02} & P_{12} & 1+P_{22} & P_{32} \\
        \ii P_{03} & \ii P_{13} & \ii P_{23} & 1+\ii P_{33} 
      \end{array}
    \right)^{-1}
    \left(
      \begin{array}{c@{\enspace}c@{\enspace}c@{\enspace}c}
        1 & 0 & 0 & 0 \\
        0 & \ii & 0 & 0 \\
        0 & 0 & -1 & 0 \\
        0 & 0 & 0 & -\ii
      \end{array}
    \right)
  \right.
  \nonumber\\
  +\left.
    \left(
      \begin{array}{c@{\enspace}c@{\enspace}c@{\enspace}c}
        1 - P_{00} & -P_{10} & -P_{20} & -P_{30} \\
        \ii P_{01} & 1+\ii P_{11} & \ii P_{21} & \ii P_{31} \\
        P_{02} & P_{12} & 1+P_{22} & P_{32} \\
        -\ii P_{03} & -\ii P_{13} & -\ii P_{23} & -\ii P_{33} 
      \end{array}
    \right)^{-1}
    \left(
      \begin{array}{c@{\enspace}c@{\enspace}c@{\enspace}c}
        1 & 0 & 0 & 0 \\
        0 & -\ii & 0 & 0 \\
        0 & 0 & -1 & 0 \\
        0 & 0 & 0 & \ii
      \end{array}
    \right)
  \right]
  \left(
    \begin{array}{c}
      \Pp_1\\
      \Pp_2\\
      \Pp_3\\
      \Pp_4
    \end{array}
  \right).
  \nonumber
\end{eqnarray}

For a symmetric walk, we substitute
\begin{equation}
    P_{03} = P_{01} = \frac{1-P_{00}-P_{02}}{2}\,,\quad
    P_{23} = P_{21} = \frac{1-P_{20}-P_{22}}{2}\,,
\end{equation}
and write 
\begin{equation}
  \eqalign{
    P_{33} = P_{11} \eqdef \pss\,,   \quad
    P_{10} = P_{30} \eqdef \psf\,,   \\
    P_{12} = P_{32} \eqdef \psb\,,   \quad
    P_{31} = P_{13} = 1 - \psf - \pss - \psb\,,
    }
\end{equation}
in terms of which the diffusion coefficient is found to be
\begin{eqnarray}
  \fl
  D_\mathrm{2SMA}
  =  D_\mathrm{NMA}
  \Bigg\{
  2 
  - 4 P_{00} 
  + 2 P_{00}^2 
  - 3 \psb 
  + 6 P_{00} \psb 
  - 3 P_{00}^2 \psb 
  + \psb^2 
  - 2 P_{00} \psb^2 
  + P_{00}^2 \psb^2 
  \nonumber\\
  + 3 \psf 
  - 6 P_{02} \psf 
  - 2 P_{00} \psf 
  + 2 P_{02} P_{00} \psf 
  - P_{00}^2 \psf 
  + 2 \psb \psf 
  + 3 P_{02} \psb \psf 
  \nonumber\\
  + 3 P_{00} \psb \psf 
  - P_{02} P_{00} \psb \psf 
  + P_{00}^2 \psb \psf 
  - 2 \psb^2 \psf 
  - 2 P_{00} \psb^2 \psf 
  + \psf^2 
  \nonumber\\
  + P_{02} \psf^2 
  + 3 P_{00} \psf^2 
  - P_{02} P_{00} \psf^2 
  - 4 \psb \psf^2 
  + 2 P_{02} \psb \psf^2 
  - 2 P_{00} \psb \psf^2 
  - 2 \psf^3 
  \nonumber\\
  + 2 P_{02} \psf^3 
  - 2 \Big[-(-1 + P_{00})^2 (-1 + \psb) + [1 + P_{02} (-3 + P_{00}) - 3 P_{00} 
  \nonumber\\
  + 2 (1 + P_{00}) \psb] \psf - 2 (-1 + P_{02}) \psf^2\Big] \pss 
  + P_{20}^2 (P_{02} - \psb) [P_{02} (-2 + \psb 
  \nonumber\\
  + \psf + 2 \pss) - 
  2 \psb (-1 + \psb + \psf + 2 \pss)]
  + P_{20} [(1 + P_{00}) \psb - P_{02} \psf] 
  \nonumber\\
  \times[3 \psb + \psf + 2 \pss + 
  P_{02} (-2 + \psb + \psf + 2 \pss)
  \nonumber\\
  - 2 \psb (\psb + \psf + 2 \pss)] 
  - P_{22}^2 \Big[2 - \psb + 3 \psf - 2 \pss 
  \nonumber\\
  - P_{00}^2 (-2 + \psb + \psf + 2 \pss) 
  - 
  \psf [\psb + 2 \psb \psf + 2 \pss 
  \nonumber\\
  - \psf (1 - 2 \psf - 4 \pss)] + 
  P_{00} [\psb (2 + 3 \psf) 
  - 4 (1 - \pss) 
  \nonumber\\
  + 
  \psf (-2 + 3 \psf + 6 \pss)]\Big] 
  - P_{22} \Big\{\psb^2 [(-1 + P_{00})^2 - 2 (1 + P_{00}) \psf] 
  \nonumber\\
  + P_{02} \psf [-6 + \psf + 6 \pss 
    - P_{00} (-2 + \psf + 2 \pss) + 
  2 \psf (\psf + 2 \pss)] + 
  \nonumber\\
  \psb \Big[\psf (3 + 3 P_{02} - 2 \psf + 2 P_{02} \psf - 4 \pss) + 
  2 (-1 + \pss) 
  \nonumber\\
  + P_{00}^2 (-2 + \psf + 2 \pss) - 
  P_{00} [-4 + P_{02} \psf + 2 \psf^2 + 4 (1 + \psf) \pss]\Big] 
  \nonumber\\
  + 
  P_{20} \Big[P_{02} 
  [4 - 2 \psb + 2 \psf - 4 \pss + 2 P_{00} (-2 + \psb + \psf + 2 \pss) 
  \nonumber\\
  - 
  3 \psf (\psb + \psf + 2 \pss)] + 
  \psb [-4 + \psb - 3 \psf + 2 \pss 
  \nonumber\\
  + 4 \psf (\psb + \psf + 2 \pss) - 
  P_{00} (-4 + 3 \psb + 3 \psf + 6 \pss)]\Big]\Big\}\Bigg\}
  \nonumber\\
  \Bigg/
  \Bigg\{
  [1 + \psb - P_{00} (1 + \psb) + P_{20} (-P_{02} + \psb) + P_{22} (-1 + P_{00} - \psf) 
  \nonumber\\
  + \psf + P_{02} \psf]
  \Big[2 (-1 + P_{00}) \psb^2 - 2 P_{02} \psf^2 
  + \psf [1 + P_{22} - P_{00} - P_{22} P_{00} 
  \nonumber\\
  + P_{02} (2 + P_{20} - 4 \pss)] +  
  2 [1 + P_{22} + P_{20} P_{02} - (1 + P_{22}) P_{00}] \pss 
  \nonumber\\
  + 
  \psb [3 + P_{22} + P_{20} P_{02} - P_{22} P_{00} - 2 (1 + P_{02}) \psf - 4 \pss 
  \nonumber\\
  + 
  P_{00} (-3 + 2 \psf + 4 \pss)]\Big]\Bigg\}\,.
\end{eqnarray}
If we further assume complete left--right symmetry, so that
\begin{equation}
  \eqalign{
    P_{33} = P_{11} = P_{31} = P_{13}
    = \frac{1 - \psf - \psb}{2}\,,
    }
\end{equation}
then the diffusion coefficient becomes 
\begin{eqnarray}
  \fl 
  D_\mathrm{2SMA}^\mathrm{s} 
  = D_\mathrm{NMA} 
  \Big[-1 + P_{22}^2 + 2 P_{22} P_{20} P_{02} + P_{20}^2 P_{02}^2 + 2 P_{00} - 2 P_{22}^2 
    P_{00} - 2 P_{22} P_{20} P_{02} P_{00} 
    \nonumber\\
    \lo
    - P_{00}^2 + P_{22}^2 P_{00}^2 + \psb - P_{22} \psb - P_{20} \psb - 3 P_{22} P_{20}
    \psb + P_{20} P_{02} \psb \nonumber\\
    \lo
    - P_{20}^2 P_{02}      
    \psb - 2 P_{00} \psb + 2 P_{22} P_{00} \psb - P_{20} P_{00} \psb 
    + P_{22} P_{20} P_{00} \psb 
    \nonumber\\
    \lo    
    + P_{20} P_{02} P_{00} \psb + P_{00}^2 \psb - P_{22} P_{00}^2 \psb - 3 \psf 
    + 3 P_{22}^2 \psf + 3 P_{02} \psf 
    \nonumber\\
    \lo    
    - 3 P_{02} \psf + P_{20} P_{02} \psf + P_{22} 
    P_{20} P_{02} \psf - P_{20} P_{02}^2 \psf + P_{00} \psf - P_{22}^2 P_{00} \psf 
    \nonumber\\
    \lo        
    - P_{02} P_{00} \psf + P_{22} P_{02} P_{00} \psf\Big]/
    \Big[-1 + P_{22}^2 + 2 P_{22} P_{20} P_{02} + P_{20}^2 P_{02}^2 + 2 P_{00} 
    \nonumber\\
    \lo        
    - 2 P_{22}^2 P_{00} -  2 P_{22} P_{20} P_{02} P_{00} - P_{00}^2 + P_{22}^2 P_{00}^2 -
    \psb - P_{22} \psb - P_{20} \psb 
    \nonumber\\
    \lo        
    -  P_{22} P_{20} \psb - P_{20} P_{02} \psb -
    P_{20}^2 P_{02} \psb 
    + 2 P_{00} \psb +  2 P_{22} P_{00} \psb + P_{20} P_{00} \psb 
    \nonumber\\
    \lo        
    + P_{22} P_{20} P_{00}
    \psb + P_{20} P_{02} P_{00} \psb - P_{00}^2 \psb - P_{22} P_{00}^2 \psb - \psf +
    P_{22}^2 \psf 
    \nonumber\\
    \lo        
    - P_{02} \psf - P_{22} P_{02} \psf - P_{20} P_{02} \psf + P_{22} P_{20}
    P_{02} \psf - P_{20} P_{02}^2 \psf + P_{00} \psf 
    \nonumber\\
    \lo        
    - P_{22}^2 P_{00} \psf + P_{02} P_{00}
    \psf + P_{22} P_{02} P_{00} \psf \Big]\,.
    \label{dcoef2spa2dqlsym}
\end{eqnarray}  
It can be checked that this equation boils down to the
expression (\ref{D1SMASq}) in the single-step memory approximation.

The validity of equations (\ref{dcoef2spa1d}),  (\ref{dcoef2spahc}) and
(\ref{dcoef2spa2dql}) has been checked by comparison with 
diffusion coefficients calculated from direct numerical simulations 
of the corresponding persistent random walk processes.

\subsection{\label{sub.2smatr}Triangular lattice}

Consider finally the triangular lattice with 6-fold symmetry. We denote the
relative directions by numbers from $0$ to $5$, following the conventions
shown in figure \ref{fig.num}(d). 

The symmetry of $\PM$ is similar to the previous subsections, so that the
structure of the problem is by now clear.
Having identified the matrix $\D$ and invariant measure $\PP$ in equation
(\ref{v0vn2SMA}), the velocity auto-correlation is found to be
\begin{eqnarray}
  \fl 
  \langle \vv_0 \cdot \vv_{n+1} \rangle = 
  3 \Big[2 \mn{n}{1, 1} + \mn{n}{1, 8} - \mn{n}{1, 15} - 2 \mn{n}{1, 22} -
  \mn{n}{1, 29} +   \mn{n}{1, 36} + \mn{n}{7, 1} + 2 \mn{n}{7, 8} 
  \nonumber\\
  +
  \mn{n}{7, 15} - \mn{n}{7, 22} -   2 \mn{n}{7, 29} - \mn{n}{7, 36} -
  \mn{n}{13, 1} + \mn{n}{13, 8} + 2 \mn{n}{13, 15} +   \mn{n}{13, 22} 
  \nonumber\\
  - \mn{n}{13, 29} - 2 \mn{n}{13, 36} - 2 \mn{n}{19, 1} - \mn{n}{19, 8} +
  \mn{n}{19, 15} + 2 \mn{n}{19, 22} + \mn{n}{19, 29} - \mn{n}{19, 36} 
  \nonumber\\
  - \mn{n}{25, 1} - 2 \mn{n}{25, 8} - \mn{n}{25, 15} + \mn{n}{25, 22} + 2
  \mn{n}{25, 29} +  \mn{n}{25, 36} + \mn{n}{31, 1} - \mn{n}{31, 8} 
    \nonumber\\
    - 2
  \mn{n}{31, 15} - \mn{n}{31, 22} +   \mn{n}{31, 29} + 2 \mn{n}{31,
    36}\Big] \Pp_{1}   \nonumber\\
  +   3 \Big[\mn{n}{1, 2} - \mn{n}{1, 9} - 2 \mn{n}{1, 16} - \mn{n}{1, 23}
  + \mn{n}{1, 30} +   2 \mn{n}{1, 31} + 2 \mn{n}{7, 2} + \mn{n}{7, 9} 
  \nonumber\\
  - \mn{n}{7, 16} - 2 \mn{n}{7, 23} -   \mn{n}{7, 30} +
  \mn{n}{7, 31} + \mn{n}{13, 2} + 2 \mn{n}{13, 9} + \mn{n}{13, 16} -
  \mn{n}{13, 23} 
  \nonumber\\
  - 2 \mn{n}{13, 30} - \mn{n}{13, 31} - \mn{n}{19, 2} +
  \mn{n}{19, 9} +   2 \mn{n}{19, 16} + \mn{n}{19, 23} - \mn{n}{19, 30} - 2
  \mn{n}{19, 31} 
  \nonumber\\
  -   2 \mn{n}{25, 2} - \mn{n}{25, 9} + \mn{n}{25, 16} + 2
  \mn{n}{25, 23} + \mn{n}{25, 30} -  \mn{n}{25, 31} - \mn{n}{31, 2} - 2
  \mn{n}{31, 9} 
  \nonumber\\
  - \mn{n}{31, 16} + \mn{n}{31, 23} +   2 \mn{n}{31, 30} +
  \mn{n}{31, 31}\Big] \Pp_{2} 
  \nonumber\\
  +   3 \Big[-\mn{n}{1, 3} - 2 \mn{n}{1, 10} -
  \mn{n}{1, 17} + \mn{n}{1, 24} + 2 \mn{n}{1, 25} +   \mn{n}{1, 32} +
  \mn{n}{7, 3} - \mn{n}{7, 10} 
  \nonumber\\
  - 2 \mn{n}{7, 17} - \mn{n}{7, 24} +
  \mn{n}{7, 25} + 2 \mn{n}{7, 32} + 2 \mn{n}{13, 3} + \mn{n}{13, 10} -
  \mn{n}{13, 17} -   2 \mn{n}{13, 24} 
  \nonumber\\
  - \mn{n}{13, 25} + \mn{n}{13, 32} +
  \mn{n}{19, 3} +   2 \mn{n}{19, 10} + \mn{n}{19, 17} - \mn{n}{19, 24} - 2
  \mn{n}{19, 25} -   \mn{n}{19, 32} 
  \nonumber\\
  - \mn{n}{25, 3} + \mn{n}{25, 10} + 2
  \mn{n}{25, 17} + \mn{n}{25, 24} -   \mn{n}{25, 25} - 2 \mn{n}{25, 32} - 2
  \mn{n}{31, 3} - \mn{n}{31, 10} 
  \nonumber\\
  +   \mn{n}{31, 17} + 2 \mn{n}{31, 24} +
  \mn{n}{31, 25} - \mn{n}{31, 32}\Big] \Pp_{3} 
  \nonumber\\
  +   3 \Big[-2 \mn{n}{1, 4} -
  \mn{n}{1, 11} + \mn{n}{1, 18} + 2 \mn{n}{1, 19} + \mn{n}{1, 26} -
  \mn{n}{1, 33} - \mn{n}{7, 4} - 2 \mn{n}{7, 11} 
  \nonumber\\
  - \mn{n}{7, 18} +
  \mn{n}{7, 19} +   2 \mn{n}{7, 26} + \mn{n}{7, 33} + \mn{n}{13, 4} -
  \mn{n}{13, 11} - 2 \mn{n}{13, 18} -   \mn{n}{13, 19} 
  \nonumber\\
  + \mn{n}{13, 26} + 2
  \mn{n}{13, 33} + 2 \mn{n}{19, 4} +   \mn{n}{19, 11} - \mn{n}{19, 18} - 2
  \mn{n}{19, 19} - \mn{n}{19, 26} + \mn{n}{19, 33} 
  \nonumber\\
  +  \mn{n}{25, 4} + 2
  \mn{n}{25, 11} + \mn{n}{25, 18} - \mn{n}{25, 19} -   2 \mn{n}{25, 26} -
  \mn{n}{25, 33} - \mn{n}{31, 4} + \mn{n}{31, 11} 
  \nonumber\\
  +   2 \mn{n}{31, 18} +
  \mn{n}{31, 19} - \mn{n}{31, 26} - 2 \mn{n}{31, 33}\Big] \Pp_{4} 
  \nonumber\\
  +   3 \Big[-\mn{n}{1, 5} + \mn{n}{1, 12} + 2 \mn{n}{1, 13} + \mn{n}{1,
    20} - \mn{n}{1, 27} -   2 \mn{n}{1, 34} - 2 \mn{n}{7, 5} 
  - \mn{n}{7, 12} 
  \nonumber\\
  + \mn{n}{7, 13} + 2 \mn{n}{7, 20} +   \mn{n}{7, 27} - \mn{n}{7, 34} -
  \mn{n}{13, 5} - 2 \mn{n}{13, 12} - \mn{n}{13, 13} +   \mn{n}{13, 20} 
  \nonumber\\
  + 2 \mn{n}{13, 27} + \mn{n}{13, 34} + \mn{n}{19, 5} - \mn{n}{19, 12} 
  -   2 \mn{n}{19, 13} - \mn{n}{19, 20} + \mn{n}{19, 27} + 2 \mn{n}{19, 34}
  \nonumber\\
  +   2 \mn{n}{25, 5} + \mn{n}{25, 12} - \mn{n}{25, 13} - 2 \mn{n}{25, 20} -
  \mn{n}{25, 27} + \mn{n}{25, 34} + \mn{n}{31, 5} + 2 \mn{n}{31, 12} 
  \nonumber\\
  +
  \mn{n}{31, 13} -   \mn{n}{31, 20} - 2 \mn{n}{31, 27} - \mn{n}{31,
    34}\Big] \Pp_{5} 
  \nonumber\\
  +   3 \Big[\mn{n}{1, 6} + 2 \mn{n}{1, 7} + \mn{n}{1, 14} - \mn{n}{1, 21} 
  - 2 \mn{n}{1, 28} -   \mn{n}{1, 35} - \mn{n}{7, 6} + \mn{n}{7, 7} 
  \nonumber\\
  + 2 \mn{n}{7, 14} + \mn{n}{7, 21} -   \mn{n}{7, 28} - 2
  \mn{n}{7, 35} - 2 \mn{n}{13, 6} - \mn{n}{13, 7} + \mn{n}{13, 14} +   2
  \mn{n}{13, 21} 
  \nonumber\\
  + \mn{n}{13, 28} - \mn{n}{13, 35} - \mn{n}{19, 6} - 2
  \mn{n}{19, 7} -  \mn{n}{19, 14} + \mn{n}{19, 21} + 2 \mn{n}{19, 28} +
  \mn{n}{19, 35} 
  \nonumber\\
  + \mn{n}{25, 6} -  \mn{n}{25, 7} - 2 \mn{n}{25, 14} -
  \mn{n}{25, 21} + \mn{n}{25, 28} +   2 \mn{n}{25, 35} + 2 \mn{n}{31, 6} +
  \mn{n}{31, 7} 
  \nonumber\\
  - \mn{n}{31, 14}   -   2 \mn{n}{31, 21} - \mn{n}{31, 28} +
  \mn{n}{31, 35}\Big] \Pp_{6}
  \end{eqnarray}
Letting $\phi = \exp(2\pi \ii/6)$, we find
\begin{eqnarray}
  \fl 
  \langle \vv_0 \cdot \vv_{n+1} \rangle  = 
  3\Big(1 \enspace 1 \enspace 1 \enspace 1 \enspace 1 \enspace 1\Big)\times
  \label{v0vn2smatr}\\
  \fl
  \left[
    \left(
      \begin{array}{c@{\enspace}c@{\enspace}c@{\enspace}c@{\enspace}c@{\enspace}c}
        \ds P_{00} & \ds P_{10} & \ds P_{20} & \ds P_{30} & \ds P_{40} &
        \ds P_{50} \\  
        \ds \phi P_{01} & \ds \phi P_{11} & \ds \phi P_{21} & \ds \phi
        P_{31} & \ds \phi P_{41} 
        & \ds \phi P_{51} \\
        \ds \phi^2 P_{02} & \ds \phi^2 P_{12} & \ds \phi^2 P_{22} & \ds
        \phi^2 P_{32} & 
        \ds \phi^2 P_{42} & \ds \phi^2 P_{52} \\  
        \ds - P_{03} & \ds - P_{13} & \ds - P_{23} & \ds - P_{33} & \ds -
        P_{33} 
        & \ds - P_{53} \\   
        \ds \phi^{-2} P_{04} & \ds \phi^{-2} P_{14} & \ds \phi^{-2} P_{24}
        & \ds \phi^{-2} P_{34} & \ds \phi^{-2} P_{44} & \ds \phi^{-2}
        P_{54} \\   
        \ds \phi^{-1} P_{05} & \ds \phi^{-1} P_{15} & \ds \phi^{-1} P_{25}
        & \ds \phi^{-1} P_{35} & \ds \phi^{-1} P_{45} & \ds \phi^{-1} P_{55}
      \end{array}
    \right)^{n}
    \left(
      \begin{array}{c@{\enspace}c@{\enspace}c@{\enspace}c@{\enspace}c@{\enspace}c}
        \ds 1 & \ds 0 & \ds 0 & \ds 0 & \ds 0 & \ds 0 \\
        \ds 0 & \ds \phi & \ds 0 & \ds 0 & \ds 0 & \ds 0 \\
        \ds 0 & \ds 0 & \ds \phi^2 & \ds 0 & \ds 0 & \ds 0 \\
        \ds 0 & \ds 0 & \ds 0 & \ds -1 & \ds 0 & \ds 0 \\
        \ds 0 & \ds 0 & \ds 0 & \ds 0 & \ds \phi^{-2} & \ds 0 \\
        \ds 0 & \ds 0 & \ds 0 & \ds 0 & \ds 0 & \ds \phi^{-1}
      \end{array}
    \right)
  \right.\nonumber\\
  \fl
  \left.
    +
    \left(
      \begin{array}{c@{\enspace}c@{\enspace}c@{\enspace}c@{\enspace}c@{\enspace}c}
        \ds P_{00} & \ds P_{10} & \ds P_{20} & \ds P_{30} & \ds P_{40} &
        \ds P_{50} \\  
        \ds \phi^{-1} P_{01} & \ds \phi^{-1} P_{11} & \ds \phi^{-1} P_{21}
        & \ds \phi^{-1} P_{31} & \ds \phi^{-1} P_{41} & \ds \phi^{-1}
        P_{51} \\ 
        \ds \phi^{-2} P_{02} & \ds \phi^{-2} P_{12} & \ds \phi^{-2} P_{22}
        & \ds \phi^{-2} P_{32} & 
        \ds \phi^{-2} P_{42} & \ds \phi^{-2} P_{52} \\  
        \ds - P_{03} & \ds - P_{13} & \ds - P_{23} & \ds - P_{33} & \ds -
        P_{33} 
        & \ds - P_{53} \\   
        \ds \phi^2 P_{04} & \ds \phi^2 P_{14} & \ds \phi^2 P_{24} & \phi^2
        \ds P_{34} & \ds \phi^2 P_{44} & \ds \phi^2 P_{54} \\  
        \ds \phi P_{05} & \ds \phi P_{15} & \ds \phi P_{25} & \phi
        \ds P_{35} & \ds \phi P_{45} & \ds \phi P_{55}
      \end{array}
    \right)^{n}
    \left(
      \begin{array}{c@{\enspace}c@{\enspace}c@{\enspace}c@{\enspace}c@{\enspace}c}
        \ds 1 & \ds 0 & \ds 0 & \ds 0 & \ds 0 & \ds 0 \\
        \ds 0 & \ds \phi^{-1} & \ds 0 & \ds 0 & \ds 0 & \ds 0 \\
        \ds 0 & \ds 0 & \ds \phi^{-2} & \ds 0 & \ds 0 & \ds 0 \\
        \ds 0 & \ds 0 & \ds 0 & \ds -1 & \ds 0 & \ds 0 \\
        \ds 0 & \ds 0 & \ds 0 & \ds 0 & \ds \phi^2 & \ds 0 \\
        \ds 0 & \ds 0 & \ds 0 & \ds 0 & \ds 0 & \ds \phi
      \end{array}
    \right)
  \right]
  \left(
    \begin{array}{c}
      \Pp_1\\
      \Pp_2\\
      \Pp_3\\
      \Pp_4\\
      \Pp_5\\
      \Pp_6
    \end{array}
  \right)
  ,
  \nonumber
\end{eqnarray}
which is again equation (\ref{v0vn2SMAgen}). The expression of the diffusion
coefficient is thus given by (\ref{dcoef2SMAgen}).

\section{\label{sec.2rev}Two-Step Memory Approximation revisited}

As seen in the previous section, the symbolic computation of 
(\ref{v0vn2SMA}) quickly becomes tricky. However, an alternative to the
above scheme can be found, provided that the walk has special symmetries.
Returning to  (\ref{2spa0}), we write  
\begin{eqnarray}
  \langle   \vv_{0} \cdot \vv_{n} \rangle
  =& \sum_{\vv_{0}} 
  \sum_{i_1,\ldots, i_n}
  \vv_{0} \cdot \RT^{i_1,\ldots,i_n}\vv_{0} 
  P(\RT^{i_1,\ldots,i_n}\vv_{0}|\RT^{i_{1},\ldots,i_{n-1}}\vv_{0}, 
  \RT^{i_{1},\ldots,i_{n-2}}\vv_{0}) \nonumber\\
  & \times \cdots\times
  P(\RT^{i_1,i_2}\vv_{0}|\RT^{i_1}\vv_{0}, \vv_{0})
  p(\vv_{0}, \RT^{i_1} \vv_{0})\,,
  \label{2spa}
\end{eqnarray}
where we introduce the compact notation $\RT^i \defeq
\R^i \T$, and sequences in the exponent denote multiple
composition: $\RT^{i_1,\dots,i_n} \defeq \RT^{i_n} \circ \RT^{i_{n-1}}
\circ \cdots \circ \RT^{i_1}$, where each $i_k$ takes values between $1$
and $z$. Note that in general $\RT^j\circ\RT^i\neq
  \RT^{i+j}$ when $\T$ is non-trivial. The 
transition probabilities $P(\RT^{j_2}\vv_{0}|\RT^{j_1}\vv_{0},
S^{j_0}\vv_{0})$ can be seen as matrix elements $\tilde \Q_{j_2 -
  j_1, j_1 - j_0}$, so that 
(\ref{2spa}) may be rewritten as
\begin{equation}
  \langle   \vv_{0} \cdot \vv_{n} \rangle
  = \sum_{\vv_{0}} \sum_{i_1,\dots,i_n} 
  \vv_{0} \cdot \RT^{i_1,\dots,i_n}\vv_{0} 
  \tilde \Q_{i_n,i_{n-1}} \cdots \tilde \Q_{i_2, i_1}
  p(\vv_{0}, \RT^{i_1}\vv_{0})\,.
  \label{2spab}
\end{equation}

We would like to rewrite this expression as a matrix product. However,
this is in general not possible, and further approximations are needed. 
Thus, assuming that the scalar product 
$\vv_{n} \cdot \vv_{0}$ factorises as 
\begin{equation}
  \RT^{i_1,\dots,i_n}\vv_{0} \cdot \vv_{0}
  =  \prod_{k=1}^n \RT^{i_k}\vv_{0} \cdot \vv_{0}\,,
  \label{factor}
\end{equation}
and defining 
\begin{equation}
  \Q_{i, j} \equiv \tilde \Q_{i, j} \RT^{j}\vv_{0} \cdot \vv_{0}\,
  = P(\RT^{j,i}\vv_{0}|\RT^{j}\vv_{0}, \vv_{0})
  \RT^{j}\vv_{0} \cdot \vv_{0}\,,
  \label{defQ}
\end{equation}
equation (\ref{2spab}) becomes
\begin{eqnarray}
  \langle   \vv_{0} \cdot \vv_{n} \rangle
  &= \sum_{\vv_{0}} \sum_{i_1,\dots,i_n} 
  \vv_{0} \cdot \RT^{i_n}\vv_{0} 
  \Q_{i_n,i_{n-1}} \cdots \Q_{i_2, i_1}\, p(\vv_{0},
  \RT^{i_1}\vv_{0})\,,\nonumber\\
  &= z \, \sum_{i_1,i_n}  \V_{i_n}^\dagger
  \Q^{n-1}_{i_n, i_1} \Pp_{i_1}
  \,,
  \label{2spac}
\end{eqnarray}
where we have introduced the vectors $\V_{i_n} \equiv \vv_{0} \cdot
\RT^{i_n}\vv_{0}$ and $\Pp_{i_1} \equiv p(\vv_{0},
\RT^{i_1}\vv_{0})$.
As can be seen, equation (\ref{2spac}) has an appropriate matrix form and
can easily be resummed over $n$ to
compute the diffusion coefficient (\ref{dcoef}).

Since $\Q$ is a $z\times z$ matrix, equation (\ref{2spac}) is much easier
to evaluate than (\ref{v0vn2SMA}). The trouble is that equation (\ref{factor})
is in general incorrect, and turns 
out to be strictly valid only for one-dimensional walks.
Nonetheless, it may also be applied to higher-dimensional walks satisfying
special
symmetry conditions. We consider the
different geometries separately in the following and discuss the conditions
under which  equation (\ref{2spac}) can be applied.  For higher-dimensional
lattices, we recover by this simpler method the results obtained earlier under
the relevant symmetry assumptions.

\subsection{\label{sub.rev1d}One-dimensional lattice}

The result (\ref{dcoef2spa1d}) follows from equation (\ref{2spac}). Indeed,
$\R^{i_n,\ldots,i_1}\vv_{0} \cdot \vv_{0} =
\R^{i_n+\cdots+i_1}\vv_{0} \cdot \vv_{0} = \pm 1$ according to 
the parity of $i_n+\cdots+i_1$, and since this is also a property of the
product $\prod_{k=1}^n \R^{i_k}\vv_{0} \cdot \vv_{0}$, we see that
equation (\ref{factor}) is valid.

The vector $\Pp_{i}$ on the right-hand side of equation (\ref{2spac}) is
\begin{equation}
  \eqalign{
    \Pp_1 = p(-v,v) = \frac{1 - P_{00}}
    {2(1 - P_{00} + P_{10})}\,,\\
    \Pp_2 = p(v,v) = \frac{P_{10}}
    {2(1 - P_{00} + P_{10})}\,.\\
    }
\end{equation}
The vector $\V_i$, on the other hand, has components
\begin{equation}
  \eqalign{
    \V_1 = -1\,,\\
    \V_2 = 1\,.
  }
\end{equation}
The matrix elements $\Q_{i,j}$ are defined according to equation
(\ref{defQ}),
\begin{eqnarray}
  \Q &= \left(
    \begin{array}{c@{\enspace}c}
      - P_{11} & P_{01}\\
      - P_{10} & P_{00}
    \end{array}
    \right)\,
    \nonumber\\
    &= \left(
      \begin{array}{c@{\enspace}c}
        P_{10} - 1 & 1 - P_{00}\\
        -P_{10} & P_{00}
      \end{array}
    \right)\,.
    \label{Q1d}
\end{eqnarray}

Considering equation (\ref{dcoef}) and plugging the above expressions into
equation (\ref{2spac}), we have
\begin{equation}
  D_\mathrm{2SMA} = D_\mathrm{NMA} \left\{1 + 
    4 \V^\dagger \left[\mathsf{I}_2 - \Q\right]^{-1} \Pp\right\}\,,
\end{equation}
and we recover equation (\ref{dcoef2spa1d}).

\subsection{\label{sub.revhc}Two-dimensional honeycomb lattice}

Consider equation (\ref{factor}) in the case of a honeycomb
lattice. The 
operation $\RT^i \vv$ is a clockwise rotation of $\vv$ by angle
$-\pi/3$ if $i=1$, $\pi/3$ if $i=2$, or $\pi$ if $i=3$. The operation
$\RT^{i_1, \dots, i_n} \vv$ is thus a rotation by angle $[2(i_1 +
\dots + i_n) - 3 n]\pi/3$, and the scalar product
\begin{equation}
  \RT^{i_1, \dots, i_n} \vv\cdot \vv = 
  \cos[2(i_1 + \dots + i_n) - 3 n]\pi/3\,.
  \label{hcfactorlhs}
\end{equation}
This expression is, however, in general different from the product
\begin{equation}
  \RT^{i_n}\vv\cdot \vv \cdots \RT^{i_1} \vv\cdot \vv
  = \prod_{k=1}^n \cos[2(i_k) - 3]\pi/3\,.
  \label{hcfactorrhs}
\end{equation}
This is so, for instance, when $n=2$ and $i_1 = i_2 = 1$, for which 
(\ref{hcfactorlhs}) yields $-1/2$, whereas (\ref{hcfactorrhs})
yields $1/4$.

There is however a special case under which the product structure that we
seek can be retrieved, as follows. 
There are a priori nine transition probabilities $P(\RT^{j,i} \vv|
\RT^{j} \vv, \vv)$. There are, however, a number of left--right
symmetries in the system which reduce the number of independent transition
probabilities to three:
\begin{equation}
    P(\RT^{1,1} \vv| \RT^{1} \vv, \vv)\,,\enspace
    P(\RT^{3,3} \vv| \RT^{3} \vv, \vv)\,,\enspace
    P(\RT^{1,2} \vv| \RT^{1} \vv, \vv)\,.
\end{equation}
In the event that the two probabilities $P(\RT^{1,2} \vv| \RT^{1}
\vv, \vv)$ and $P(\RT^{1,1} \vv| \RT^{1} \vv, \vv)$ are equal,
\begin{equation}
    P(\RT^{1,2} \vv| \RT^{1} \vv, \vv) = 
    P(\RT^{1,1} \vv| \RT^{1} \vv, \vv) \equiv \pss\,,
    \label{Hc2SMAs}
\end{equation}
which is to say that forward--left and right scatterings are treated as
identical events, and the number of independent parameters reduces to two,
which we take to be $P_{00}$ and $P_{ss}$.

In this case, the expression of the diffusion coefficient can be
obtained in a way similar to equation (\ref{dcoef2spa1d}) for the
one-dimensional lattice. This is so because
\begin{eqnarray}
  \RT^{i_1 \dots, i_{n-1},1} \vv\cdot \vv &=& 
  \cos\left\{[2(i_1 + \dots + i_{n-1} + 1) - 3
    n]\frac{\pi}{3}\right\}\,\nonumber\\ 
  &=& \cos \frac{\pi}{3}\cos\left\{[2(i_1 + \dots + i_{n-1}) - 3
    (n-1)]\frac{\pi}{3}\right\} \nonumber\\ 
  &&+ \sin \frac{\pi}{3}\sin\left\{[2(i_1 + \dots + i_{n-1}) - 3
    (n-1)]\frac{\pi}{3}\right\} \,,
  \label{hbineq1}\\
  \RT^{i_1 \dots, i_{n-1}, 2} \vv\cdot \vv &=& 
  \cos\left\{[2(i_1 + \dots + i_{n-1} + 2) - 3
    n]\frac{\pi}{3}\right\}\,\nonumber\\ 
  &=& \cos \frac{\pi}{3}\cos\left\{[2(i_1 + \dots + i_{n-1}) - 3
    (n-1)]\frac{\pi}{3}\right\} \nonumber\\ 
  &&- \sin \frac{\pi}{3}\sin\left\{[2(i_1 + \dots + i_{n-1}) - 3
    (n-1)]\frac{\pi}{3}\right\} \,,
  \label{hbineq2}\\
  \RT^{i_1 \dots, i_{n-1}, 3} \vv\cdot \vv &=& 
  \cos\left\{[2(i_1 + \dots + i_{n-1} + 3) - 3
    n]\frac{\pi}{3}\right\}\,\nonumber\\ 
  &=& -\cos\left\{[2(i_1 + \dots + i_{n-1}) - 3
    (n-1)]\frac{\pi}{3}\right\} 
  \label{hbineq3}\,.
\end{eqnarray}
Thus, given the symmetry between forward--left and right scatterings, the
two sine contributions in equations (\ref{hbineq1}) and (\ref{hbineq2})
cancel, whereas the cosines add up to $1$:
\begin{eqnarray}
  \RT^{i_1 \dots, i_{n-1},1} \vv\cdot \vv 
  + \RT^{i_1 \dots, i_{n-1},2} \vv\cdot \vv 
  &= \RT^{i_1 \dots, i_{n-1}} \vv\cdot \vv\,,\nonumber\\
  &= \RT^{i_1 \dots, i_{n-1}} \vv\cdot \vv
  (\RT^{1} \vv\cdot \vv + \RT^{2} \vv\cdot \vv) \,,
  \\
  \RT^{i_1 \dots, i_{n-1},3} \vv\cdot \vv 
  =  -\RT^{i_1 \dots, i_{n-1}} \vv\cdot \vv
  &= \RT^{i_1 \dots, i_{n-1}} \vv\cdot \vv\,
  \RT^{3} \vv\cdot \vv\,.
\end{eqnarray}
We therefore retrieve an effective product structure, as in equation
(\ref{factor}), and can compute the diffusion coefficient using equation
(\ref{2spac}), with
\begin{equation}
  \PP =
  \frac{1}{3(2 - 2 \pss - 
    P_{00})} 
    \left( 
      \begin{array}{c}
        \frac{1}{2}(1 - P_{00}) \\ 
        1 - 2 \pss 
      \end{array}
    \right)\,,
\end{equation} 
\begin{equation}
  \V =
  \left( 
    \begin{array}{c}
      1\\-1
    \end{array}
  \right)\,,
\end{equation}
and
\begin{equation}
  \Q  = \left(
      \begin{array}{c@{\enspace}c}
        \pss 
        & 1/2 P_{00} - 1/2 \\ 
        1 - 2 \pss & 
        - P_{00} 
      \end{array}
    \right)\,.
\end{equation}
We obtain the expression of the diffusion coefficient for the symmetric
[in the sense of equation (\ref{Hc2SMAs})] two-step memory
approximation on the honeycomb lattice:
\begin{equation}
  D_\mathrm{2SMA}^\mathrm{s} 
  = D_\mathrm{NMA}
  \frac{3 (1 - P_{00})} 
  {(3 - 4 \pss 
    + P_{00})} 
  \frac{(2 \pss 
    + P_{00} )} 
  {(2 - 2 \pss 
    - P_{00})}\,. 
  \label{dcoef2spahcsymrev}
\end{equation}
This is equation (\ref{dcoef2spahcsym}).

\subsection{\label{sub.revsq}Two-dimensional square lattice}

For the two-dimensional square lattice, 
recall that $\T$ is the identity and the operation $\RT^i
\vv = \R^i \vv$ is a anticlockwise rotation of $\vv$ by angle
$i \pi/2$, with 
$i=0,\dots,3$.
The operation 
$\RT^{i_1, \dots, i_n} \vv$ is thus a rotation by angle $(i_1 +
\cdots + i_n)\pi/2$. Equations similar to (\ref{hbineq1})--(\ref{hbineq3})
hold\footnote{
Note that, in general, we have the decomposition
\begin{equation*}
  \R^{i_1 \ldots, i_{n}} \vv\cdot \vv 
  = \sum_{\omega_1,\ldots,\omega_n\in\{0,1\}} 
  \mathrm{sgn}(\omega_1,\ldots,\omega_n)
  \R^{i_1} \vv\cdot \vv_{\omega_1} \cdots   \R^{i_n} \vv\cdot
  \vv_{\omega_n} \,,
\end{equation*}
where we introduced the notation $\vv_{\omega} = \vv$ if $\omega=0$
and $\vv_{\omega} = \vv_\perp$ if $\omega=1$, and the function 
$\mathrm{sgn}(\omega_1,\dots,\omega_n) = \pm 1$, depending on the sequence 
$\omega_1,\dots,\omega_n$. Equation (\ref{2spac}) would then be replaced by
a more complicated expression involving the mixed products of two matrices
$P(\R^{j+i} \vv | \R^{j} \vv, \vv) \R^{j} \vv \cdot \vv$ and
$P(\R^{j+i} \vv | \R^{j} \vv, \vv) \R^{j}\vv \cdot
\vv_\perp$.  
}: 
\begin{eqnarray}
  \R^{i_1 \dots, i_{n}} \vv\cdot \vv 
  &=& 
  \cos\left[(i_1 + \cdots + i_{n})\frac{\pi}{2}\right]\,,
  \nonumber\\ 
  &=& \cos \frac{i_n\pi}{2}
  \cos\left[(i_1 + \cdots + i_{n-1})\frac{\pi}{2}\right]
  \nonumber\\ 
  &&- \sin \frac{i_n\pi}{2}
  \sin\left[(i_1 + \cdots + i_{n-1})\frac{\pi}{2}\right]\,,
  \nonumber\\
  &=& \R^{i_1 \ldots, i_{n-1}} \vv\cdot \vv \,
  \R^{i_{n}} \vv\cdot \vv 
  \nonumber\\ 
  &&- (\delta_{i_n,1} - \delta_{i_n,3})
  \sin\left[(i_1 + \cdots + i_{n-1})\frac{\pi}{2}\right]\,.
  \label{qlineq1}
\end{eqnarray}
The last term drops out provided 
\begin{equation}
  P(\R^{i+1} \vv| \R^{i} \vv, \vv) = P(\R^{i+3} \vv| \R^{i}
  \vv, \vv) \,.
  \label{qlcond1}
\end{equation}
Under the additional assumption that
\begin{equation}
  P(\R^{1+i} \vv| \R^{1} \vv, \vv) = P(\R^{3+i} \vv| \R^{3}
  \vv, \vv) \,,
  \label{qlcond2}
\end{equation}
we retrieve an effective factorisation similar to that
postulated in  (\ref{factor}), and we can then use 
(\ref{2spac})
to obtain the corresponding diffusion coefficient. We refer to equations
(\ref{qlcond1}) and (\ref{qlcond2}) as defining a complete left--right
symmetry. In this case, the invariant distribution is the solution of 
\begin{eqnarray}
  p(\vv, \R^i \vv) 
  = 
  \sum_j P(\R^{i+j} \vv| \R^{j} \vv, \vv)
  p(\R^i \vv, \R^j \vv)\,,\\
  \sum_i p(\vv, \R^i \vv) = \frac{1}{4}\,.
\end{eqnarray}
We solve these equations for $p(\vv, \vv)$ and $p(\vv, -\vv)$,
identifying $p(\vv, \R^1 \vv)$ and $p(\vv, \R^3 \vv)$, and
define
\begin{equation}
  \V = 
  \left( 
    \begin{array}{c}
      -1\\
      1
    \end{array}
  \right)\,,
  \enspace
  \PP = 
  \left( 
    \begin{array}{c}
      p(\vv, -\vv)\\
      p(\vv, \vv)
    \end{array}
  \right)\,,
\end{equation} 
with the transition matrix
\begin{equation}
  \Q = 
    \left( 
      \begin{array}{c@{\enspace}c}
        -P(\vv| -\vv, \vv)&
        P(- \vv| \vv, \vv)\\
        -P(-\vv| -\vv, \vv)&
        P(\vv| \vv, \vv)\\
      \end{array}
    \right) 
    = 
    \left( 
      \begin{array}{c@{\enspace}c}
        - P_{22}&
        P_{02}\\
        - P_{20}&
        P_{00}\\
      \end{array}
    \right)\,.
\end{equation} 
The resulting expression of the diffusion coefficient is identical to
 (\ref{dcoef2spa2dqlsym}).

The validity of this expression extends to $d$-dimensional orthogonal
lattices under the symmetry assumptions (\ref{qlcond1})--(\ref{qlcond2}). 

\section{\label{sec.con}Conclusions}

We have shown that it is possible to find exact results for the diffusion
coefficient of persistent random walks with two-step memory on regular
lattices, by finding the matrix elements which give the velocity
auto-correlation function and then resumming then.

We have applied the results
obtained here to approximate the diffusion coefficients of certain
periodic billiard tables in \cite{GilbertSanders09}.

The extension to lattice random walks with longer memory is possible, albeit
difficult for obvious technical reasons.  
Finally, we remark that the extension to lattices in three dimensions is not
direct, since in that case, additional information must be specified in order to
uniquely define relative directions \cite{La97}. 

\ack
The authors thank Hern\'an Larralde 
for helpful discussions. This research benefitted from
the joint support of FNRS (Belgium) and CONACYT (Mexico) through a
bilateral collaboration project. The work of TG is financially supported by the
Belgian Federal
Government under the Inter-university Attraction Pole project NOSY
P06/02. TG is financially supported by the Fonds de la Recherche
Scientifique F.R.S.-FNRS.  DPS acknowledges financial support from
DGAPA-UNAM project  IN105209, and the hospitality of the Universit\'e Libre de
Bruxelles,
where 
most of this work was carried out.  TG acknowledges
the hospitality of the Weizmann Institute of Science, where part of this work
was completed.

\begin{appendix}

\section{\label{app.hc} 2SMA on a honeycomb lattice}

In analogy to equation (\ref{mnab}), we may write
\begin{equation}
  \PM^n \equiv 
  \left(
\begin{array}{c@{\enspace}c@{\enspace}c@{\enspace}c@{\enspace}c@{\enspace}c@{
\enspace}c@{\enspace}c@{\enspace}c}
      \an{n}{1} & \an{n}{2}& \an{n}{3} & \an{n}{4} &
      \an{n}{5} & \an{n}{6}& \an{n}{7} & \an{n}{8} & \an{n}{9} \\
      \cn{n}{9} & \cn{n}{7}& \cn{n}{8} & \cn{n}{3} &
      \cn{n}{1} & \cn{n}{2}& \cn{n}{6} & \cn{n}{4} & \cn{n}{5} \\
      \bn{n}{5} & \bn{n}{6}& \bn{n}{4} & \bn{n}{8} &
      \bn{n}{9} & \bn{n}{7}& \bn{n}{2} & \bn{n}{3} & \bn{n}{1} \\
      \bn{n}{1} & \bn{n}{2}& \bn{n}{3} & \bn{n}{4} &
      \bn{n}{5} & \bn{n}{6}& \bn{n}{7} & \bn{n}{8} & \bn{n}{9} \\
      \an{n}{9} & \an{n}{7}& \an{n}{8} & \an{n}{3} &
      \an{n}{1} & \an{n}{2}& \an{n}{6} & \an{n}{4} & \an{n}{5} \\
      \cn{n}{5} & \cn{n}{6}& \cn{n}{4} & \cn{n}{8} &
      \cn{n}{9} & \cn{n}{7}& \cn{n}{2} & \cn{n}{3} & \cn{n}{1} \\
      \cn{n}{1} & \cn{n}{2}& \cn{n}{3} & \cn{n}{4} &
      \cn{n}{5} & \cn{n}{6}& \cn{n}{7} & \cn{n}{8} & \cn{n}{9} \\
      \bn{n}{9} & \bn{n}{7}& \bn{n}{8} & \bn{n}{3} &
      \bn{n}{1} & \bn{n}{2}& \bn{n}{6} & \bn{n}{4} & \bn{n}{5} \\
      \an{n}{5} & \an{n}{6}& \an{n}{4} & \an{n}{8} &
      \an{n}{9} & \an{n}{7}& \an{n}{2} & \an{n}{3} & \an{n}{1} 
    \end{array}
  \right)\,.
  \label{mnabc}
\end{equation}

We have the three sets of equations
\begin{eqnarray}
  \fl
  \left(
    \begin{array}{c@{\enspace}c@{\enspace}c}
      \an{n}{1}& \an{n}{5} & \an{n}{9}\\
      \cn{n}{1}& \cn{n}{5} & \cn{n}{9}\\
      \bn{n}{1}& \bn{n}{5} & \bn{n}{9}
    \end{array}
  \right)
  = 
  \left(
    \begin{array}{c@{\enspace}c@{\enspace}c}
      P_{00}& P_{10}& P_{20} \\
      P_{01}& P_{11}& P_{21} \\
      P_{02}& P_{12}& P_{22} 
    \end{array}
  \right)
  \left(
    \begin{array}{c@{\enspace}c@{\enspace}c}
      \an{n-1}{1}& \an{n-1}{5} & \an{n-1}{9}\\
      \cn{n-1}{9}& \cn{n-1}{1} & \cn{n-1}{5}\\
      \bn{n-1}{5}& \bn{n-1}{9} & \bn{n-1}{1}
    \end{array}
  \right)\,,
  \label{abc159}\\
  \fl
  \left(
    \begin{array}{c@{\enspace}c@{\enspace}c}
      \an{n}{2}& \an{n}{6} & \an{n}{7}\\
      \cn{n}{2}& \cn{n}{6} & \cn{n}{7}\\
      \bn{n}{2}& \bn{n}{6} & \bn{n}{7}
    \end{array}
  \right)
  = 
  \left(
    \begin{array}{c@{\enspace}c@{\enspace}c}
      P_{00}& P_{10}& P_{20} \\
      P_{01}& P_{11}& P_{21} \\
      P_{02}& P_{12}& P_{22} 
    \end{array}
  \right)
  \left(
    \begin{array}{c@{\enspace}c@{\enspace}c}
      \an{n-1}{2}& \an{n-1}{6} & \an{n-1}{7}\\
      \cn{n-1}{7}& \cn{n-1}{2} & \cn{n-1}{6}\\
      \bn{n-1}{6}& \bn{n-1}{7} & \bn{n-1}{2}
    \end{array}
  \right)\,,
  \label{abc267}\\
  \fl
  \left(
    \begin{array}{c@{\enspace}c@{\enspace}c}
      \an{n}{3}& \an{n}{4} & \an{n}{8}\\
      \cn{n}{3}& \cn{n}{4} & \cn{n}{8}\\
      \bn{n}{3}& \bn{n}{4} & \bn{n}{8}
    \end{array}
  \right)
  = 
  \left(
    \begin{array}{c@{\enspace}c@{\enspace}c}
      P_{00}& P_{10}& P_{20} \\
      P_{01}& P_{11}& P_{21} \\
      P_{02}& P_{12}& P_{22} 
    \end{array}
  \right)
  \left(
    \begin{array}{c@{\enspace}c@{\enspace}c}
      \an{n-1}{3}& \an{n-1}{4} & \an{n-1}{8}\\
      \cn{n-1}{8}& \cn{n-1}{3} & \cn{n-1}{4}\\
      \bn{n-1}{4}& \bn{n-1}{8} & \bn{n-1}{3}
    \end{array}
  \right)\,.
  \label{abc348}
\end{eqnarray}
Proceeding with our analogy, we seek linear combinations of the
elements in the rows of the matrices on the left-hand side of the above
equations,
so as to obtain a single matrix equation similar to equation
(\ref{recursionab}). Considering the elements in equation (\ref{abc159}),
we write 
\begin{eqnarray}
  \fl
  \left(
    \begin{array}{c}
      \an{n}{1} + \phi \an{n}{5} + \phi^2 \an{n}{9}\\
      \phi \cn{n}{1} + \phi^2 \cn{n}{5} + \cn{n}{9}\\
      \phi^2 \bn{n}{1} + \bn{n}{5} + \phi \bn{n}{9}
    \end{array}
  \right)
  = 
  \nonumber\\
  \left(
    \begin{array}{c@{\enspace}c@{\enspace}c}
      P_{00}& P_{10}& P_{20} \\
      \phi P_{01}& \phi P_{11}& \phi P_{21} \\
      \phi^2 P_{02}& \phi^2 P_{12}& \phi^2 P_{22} 
    \end{array}
  \right)
  \left(
    \begin{array}{c}
      \an{n-1}{1} + \phi \an{n-1}{5} + \phi^2 \an{n-1}{9}\\
      \phi \cn{n-1}{1} + \phi^2 \cn{n-1}{5} + \cn{n-1}{9}\\
      \phi^2 \bn{n-1}{1} + \bn{n-1}{5} + \phi \bn{n-1}{9}
    \end{array}
  \right)\,.
  \label{sumabc159}
\end{eqnarray}
Comparing with equation (\ref{abc159}), we infer 
\begin{equation}
    \phi^3 = 1
  \Leftrightarrow
  \left\{
    \begin{array}{l}
      \phi = 1\,,\\
      \phi = \exp(2\ii\pi/3) = -\frac{1 - \ii\sqrt{3}}{2} \,,\\
      \phi = \exp(-2\ii\pi/3) = -\frac{1 + \ii\sqrt{3}}{2} \,.
    \end{array}
  \right.
  \label{solphi}
\end{equation}
Applying the same procedure to equations (\ref{abc267})--(\ref{abc348}),
we obtain 
\begin{eqnarray}
  \fl
  \left(
    \begin{array}{c}
      \phi \an{n}{2} + \phi^2 \an{n}{6} + \an{n}{7}\\
      \phi^2 \cn{n}{2} + \cn{n}{6} + \phi \cn{n}{7}\\
      \bn{n}{2} + \phi \bn{n}{6} + \phi^2 \bn{n}{7}
    \end{array}
  \right)
  = 
  \nonumber\\
  \left(
    \begin{array}{c@{\enspace}c@{\enspace}c}
      P_{00}& P_{10}& P_{20} \\
      \phi P_{01}& \phi P_{11}& \phi P_{21} \\
      \phi^2 P_{02}& \phi^2 P_{12}& \phi^2 P_{22} 
    \end{array}
  \right)
  \left(
    \begin{array}{c}
      \phi \an{n-1}{2} + \phi^2 \an{n-1}{6} + \an{n-1}{7}\\
      \phi^2 \cn{n-1}{2} + \cn{n-1}{6} + \phi \cn{n-1}{7}\\
      \bn{n-1}{2} + \phi \bn{n-1}{6} + \phi^2 \bn{n-1}{7}
    \end{array}
  \right)\,.
  \label{sumabc267}\\
  \fl
  \left(
    \begin{array}{c}
      \phi^2 \an{n}{3} + \an{n}{4} + \phi \an{n}{8}\\
      \cn{n}{3} + \phi \cn{n}{4} + \phi^2 \cn{n}{8}\\
      \phi \bn{n}{3} + \phi^ 2 \bn{n}{4} + \bn{n}{8}
    \end{array}
  \right)
  = 
  \nonumber\\
  \left(
    \begin{array}{c@{\enspace}c@{\enspace}c}
      P_{00}& P_{10}& P_{20} \\
      \phi P_{01}& \phi P_{11}& \phi P_{21} \\
      \phi^2 P_{02}& \phi^2 P_{12}& \phi^2 P_{22} 
    \end{array}
  \right)
  \left(
    \begin{array}{c}
      \phi^2 \an{n-1}{3} + \an{n-1}{4} + \phi \an{n-1}{8}\\
      \cn{n-1}{3} + \phi \cn{n-1}{4} + \phi^2 \cn{n-1}{8}\\
      \phi \bn{n-1}{3} + \phi^ 2 \bn{n-1}{4} + \bn{n-1}{8}
    \end{array}
  \right)\,.
  \label{sumabc348}
\end{eqnarray}

The system of equations (\ref{sumabc159}), (\ref{sumabc267}),
(\ref{sumabc348}) reduces to the single recursive matrix equation  
\begin{eqnarray}
  \fl
  \left(
    \begin{array}{c@{\enspace}c@{\enspace}c}
      \an{n}{1} + \phi \an{n}{5} + \phi^2 \an{n}{9}&
      \phi \an{n}{2} + \phi^2 \an{n}{6} +  \an{n}{7}&
      \phi^2 \an{n}{3} + \an{n}{4} + \phi \an{n}{8}\\
      \phi \cn{n}{1} + \phi^2 \cn{n}{5} + \cn{n}{9}&
      \phi^2 \cn{n}{2} + \cn{n}{6} + \phi \cn{n}{7}&
      \cn{n}{3} + \phi \cn{n}{4} + \phi^2 \cn{n}{8}\\
      \phi^2 \bn{n}{1} + \bn{n}{5} + \phi \bn{n}{9}&
      \bn{n}{2} + \phi \bn{n}{6} + \phi^2 \bn{n}{7}&
      \phi \bn{n}{3} + \phi^2 \bn{n}{4} + \bn{n}{8}
    \end{array}
  \right)
  \nonumber\\
  = 
  \left(
    \begin{array}{c@{\enspace}c@{\enspace}c}
      P_{00}& P_{10}& P_{20} \\
      \phi P_{01}& \phi P_{11}& \phi P_{21} \\
       \phi^2 P_{02}& \phi^2 P_{12}& \phi^2 P_{22} 
    \end{array}
  \right)\nonumber\\
  \times \left(
    \begin{array}{c@{\enspace}c@{\enspace}c}
      \an{n-1}{1} + \phi \an{n-1}{5} + \phi^2 \an{n-1}{9}&
      \dots&
      \phi^2 \an{n-1}{3} + \an{n-1}{4} + \phi \an{n-1}{8}\\
      \phi \cn{n-1}{1} + \phi^2 \cn{n-1}{5} + \cn{n-1}{9}&
      \dots&
      \cn{n-1}{3} + \phi \cn{n-1}{4} + \phi^2 \cn{n-1}{8}\\
      \phi^2 \bn{n-1}{1} + \bn{n-1}{5} + \phi \bn{n-1}{9}&
      \dots&
      \phi \bn{n-1}{3} + \phi^2 \bn{n-1}{4} + \bn{n-1}{8}
    \end{array}
  \right)\,,
  \nonumber\\
  = 
  \left(
    \begin{array}{c@{\enspace}c@{\enspace}c}
      P_{00}& P_{10}& P_{20} \\
      \phi P_{01}& \phi P_{11}& \phi P_{21} \\
      \phi^2 P_{02}& \phi^2 P_{12}& \phi^2 P_{22} 
    \end{array}
  \right)^{n-1}
  \left(
    \begin{array}{c@{\enspace}c@{\enspace}c}
      P_{00}& \phi P_{10}& \phi^2 P_{20} \\
      \phi P_{01}& \phi^2 P_{11}& P_{21} \\
      \phi^2 P_{02}& P_{12}& \phi P_{22} 
    \end{array}
  \right)
  \,,
  \nonumber\\
  = 
  \left(
    \begin{array}{c@{\enspace}c@{\enspace}c}
      P_{00}& P_{10}& P_{20} \\
      \phi P_{01}& \phi P_{11}& \phi P_{21} \\
      \phi^2 P_{02}& \phi^2 P_{12}& \phi^2 P_{22} 
    \end{array}
  \right)^{n}
  \left(
    \begin{array}{c@{\enspace}c@{\enspace}c}
      1 & 0 & 0 \\
      0 & \phi & 0 \\
      0 & 0 & \phi^2
    \end{array}
  \right)  \,.
  \label{recursionabcapp}
\end{eqnarray}

\section{\label{app.sq} 2SMA on a square lattice}

In analogy to equations (\ref{abc159})--(\ref{abc348}), there are 64
different entries among the 256 elements of $\PM^n$, which can be obtained
through the set of equations 
\begin{eqnarray}
  \fl
  \left(
    \begin{array}{c@{\enspace}c@{\enspace}c@{\enspace}c}
      \an{n}{1}& \an{n}{6} & \an{n}{11} & \an{n}{16}\\
      \dn{n}{1}& \dn{n}{6} & \dn{n}{11} & \dn{n}{16}\\
      \cn{n}{1}& \cn{n}{6} & \cn{n}{11} & \cn{n}{16}\\
      \bn{n}{1}& \bn{n}{6} & \bn{n}{11} & \bn{n}{16}
    \end{array}
  \right)
  = 
  \left(
    \begin{array}{c@{\enspace}c@{\enspace}c@{\enspace}c}
      P_{00} & P_{10} & P_{20} & P_{30} \\
      P_{01} & P_{11} & P_{21} & P_{31} \\
      P_{02} & P_{12} & P_{22} & P_{32} \\
      P_{03} & P_{13} & P_{23} & P_{33} 
    \end{array}
  \right)
  \left(
    \begin{array}{c@{\enspace}c@{\enspace}c@{\enspace}c}
      \an{n-1}{1}& \an{n-1}{6} & \an{n-1}{11} & \an{n-1}{16}\\
      \dn{n-1}{16}& \dn{n-1}{1} & \dn{n-1}{6} & \dn{n-1}{11}\\
      \cn{n-1}{11}& \cn{n-1}{16} & \cn{n-1}{1} & \cn{n-1}{6}\\
      \bn{n-1}{6}& \bn{n-1}{11} & \bn{n-1}{16} & \bn{n-1}{1}
    \end{array}
  \right),
  \label{abcd1-6-11-16}\\
  \fl
  \left(
    \begin{array}{c@{\enspace}c@{\enspace}c@{\enspace}c}
      \an{n}{2}& \an{n}{7} & \an{n}{12} & \an{n}{13}\\
      \dn{n}{2}& \dn{n}{7} & \dn{n}{12} & \dn{n}{13}\\
      \cn{n}{2}& \cn{n}{7} & \cn{n}{12} & \cn{n}{13}\\
      \bn{n}{2}& \bn{n}{7} & \bn{n}{12} & \bn{n}{13}
    \end{array}
  \right)
  = 
  \left(
    \begin{array}{c@{\enspace}c@{\enspace}c@{\enspace}c}
      P_{00} & P_{10} & P_{20} & P_{30} \\
      P_{01} & P_{11} & P_{21} & P_{31} \\
      P_{02} & P_{12} & P_{22} & P_{32} \\
      P_{03} & P_{13} & P_{23} & P_{33} 
    \end{array}
  \right)
  \left(
    \begin{array}{c@{\enspace}c@{\enspace}c@{\enspace}c}
      \an{n-1}{2}& \an{n-1}{7} & \an{n-1}{12} & \an{n-1}{13}\\
      \dn{n-1}{13}& \dn{n-1}{2} & \dn{n-1}{7} & \dn{n-1}{12}\\
      \cn{n-1}{12}& \cn{n-1}{13} & \cn{n-1}{2} & \cn{n-1}{7}\\
      \bn{n-1}{7}& \bn{n-1}{12} & \bn{n-1}{13} & \bn{n-1}{2}
    \end{array}
  \right),
  \label{abcd2-7-12-13}\\
  \fl
  \left(
    \begin{array}{c@{\enspace}c@{\enspace}c@{\enspace}c}
      \an{n}{3}& \an{n}{8} & \an{n}{9} & \an{n}{14}\\
      \dn{n}{3}& \dn{n}{8} & \dn{n}{9} & \dn{n}{14}\\
      \cn{n}{3}& \cn{n}{8} & \cn{n}{9} & \cn{n}{14}\\
      \bn{n}{3}& \bn{n}{8} & \bn{n}{9} & \bn{n}{14}
    \end{array}
  \right)
  = 
  \left(
    \begin{array}{c@{\enspace}c@{\enspace}c@{\enspace}c}
      P_{00} & P_{10} & P_{20} & P_{30} \\
      P_{01} & P_{11} & P_{21} & P_{31} \\
      P_{02} & P_{12} & P_{22} & P_{32} \\
      P_{03} & P_{13} & P_{23} & P_{33} 
    \end{array}
  \right)
  \left(
    \begin{array}{c@{\enspace}c@{\enspace}c@{\enspace}c}
      \an{n-1}{3}& \an{n-1}{8} & \an{n-1}{9} & \an{n-1}{14}\\
      \dn{n-1}{14}& \dn{n-1}{3} & \dn{n-1}{8} & \dn{n-1}{9}\\
      \cn{n-1}{9}& \cn{n-1}{14} & \cn{n-1}{3} & \cn{n-1}{8}\\
      \bn{n-1}{8}& \bn{n-1}{9} & \bn{n-1}{14} & \bn{n-1}{3}
    \end{array}
  \right),
  \label{abcd3-8-9-14}\\
  \fl
  \left(
    \begin{array}{c@{\enspace}c@{\enspace}c@{\enspace}c}
      \an{n}{4}& \an{n}{5} & \an{n}{10} & \an{n}{15}\\
      \dn{n}{4}& \dn{n}{5} & \dn{n}{10} & \dn{n}{15}\\
      \cn{n}{4}& \cn{n}{5} & \cn{n}{10} & \cn{n}{15}\\
      \bn{n}{4}& \bn{n}{5} & \bn{n}{10} & \bn{n}{15}
    \end{array}
  \right)
  = 
  \left(
    \begin{array}{c@{\enspace}c@{\enspace}c@{\enspace}c}
      P_{00} & P_{10} & P_{20} & P_{30} \\
      P_{01} & P_{11} & P_{21} & P_{31} \\
      P_{02} & P_{12} & P_{22} & P_{32} \\
      P_{03} & P_{13} & P_{23} & P_{33} 
    \end{array}
  \right)
  \left(
    \begin{array}{c@{\enspace}c@{\enspace}c@{\enspace}c}
      \an{n-1}{4}& \an{n-1}{5} & \an{n-1}{10} & \an{n-1}{15}\\
      \dn{n-1}{15}& \dn{n-1}{4} & \dn{n-1}{5} & \dn{n-1}{10}\\
      \cn{n-1}{10}& \cn{n-1}{15} & \cn{n-1}{4} & \cn{n-1}{5}\\
      \bn{n-1}{5}& \bn{n-1}{10} & \bn{n-1}{15} & \bn{n-1}{4}
    \end{array}
  \right).
  \label{abcd4-5-10-15}
\end{eqnarray}
Combining these quantities, we let
\begin{eqnarray*}
  \An{n}{1}{k} \equiv 
  \an{n}{1} + \phi_k \an{n}{6} + \phi_k^2 \an{n}{11} + \phi_k^3
  \an{n}{16}\,,\\ 
  \An{n}{5}{k} \equiv 
  \phi_k^3 \an{n}{4} + \an{n}{5} + \phi_k \an{n}{10} + \phi_k^2
  \an{n}{15}\,,\\ 
  \An{n}{9}{k} \equiv 
  \phi_k^2 \an{n}{3} + \phi_k^3 \an{n}{8} + \an{n}{9} + \phi_k
  \an{n}{14}\,,\\ 
  \An{n}{13}{k} \equiv 
  \phi_k \an{n}{2} + \phi_k^2 \an{n}{7} + \phi_k^3 \an{n}{12} +
  \an{n}{13}\,,\\  
  \Bn{n}{2}{k} \equiv 
  \bn{n}{2} + \phi_k \bn{n}{7} + \phi_k^2 \bn{n}{12} + \phi_k^3
  \bn{n}{13}\,,\\ 
  \Bn{n}{6}{k} \equiv  
  \phi_k^3 \bn{n}{1} + \bn{n}{6} + \phi_k \bn{n}{11} + \phi_k^2
  \bn{n}{16}\,,\\
  \Bn{n}{10}{k} \equiv 
  \phi_k^2 \bn{n}{4} + \phi_k^3 \bn{n}{5} + \bn{n}{10} + \phi_k
  \bn{n}{15}\,, \\  
  \Bn{n}{14}{k} \equiv 
  \phi_k \bn{n}{3} + \phi_k^2 \bn{n}{8} + \phi_k^3 \bn{n}{9} +
  \bn{n}{14}\,,\\
  \Cn{n}{3}{k} \equiv 
  \cn{n}{3} + \phi_k \cn{n}{8} + \phi_k^2 \cn{n}{9} + \phi_k^3
  \cn{n}{14}\,,\\ 
  \Cn{n}{7}{k} \equiv 
  \phi_k^3 \cn{n}{2} + \cn{n}{7} + \phi_k \cn{n}{12} + \phi_k^2
  \cn{n}{13}\,,\\ 
  \Cn{n}{11}{k} \equiv 
  \phi_k^2 \cn{n}{1} + \phi_k^3 \cn{n}{6} + \cn{n}{11} + \phi_k
  \cn{n}{16}\,,\\ 
  \Cn{n}{15}{k} \equiv 
  \phi_k \cn{n}{4} + \phi_k^2 \cn{n}{5} + \phi_k^3 \cn{n}{10} +
  \cn{n}{15}\,, \\ 
  \Dn{n}{4}{k} \equiv 
  \dn{n}{4} + \phi_k \dn{n}{5} +\phi_k^2 \dn{n}{10} + \phi_k^3
  \dn{n}{15}\,,\\ 
  \Dn{n}{8}{k} \equiv 
  \phi_k^3 \dn{n}{3} + \dn{n}{8} + \phi_k \dn{n}{9} + \phi_k^2
  \dn{n}{14}\,, \\
  \Dn{n}{12}{k} \equiv 
  \phi_k^2 \dn{n}{2} + \phi_k^3 \dn{n}{7} + \dn{n}{12} + \phi_k
  \dn{n}{13}\,,\\ 
  \Dn{n}{16}{k} \equiv 
  \phi_k \dn{n}{1} + \phi_k^2 \dn{n}{6} + \phi_k^3 \dn{n}{11} +
  \dn{n}{16}\,,
\end{eqnarray*}
in terms of which we have
\begin{eqnarray}
  \fl
  \left(
    \begin{array}{c@{\enspace}c@{\enspace}c@{\enspace}c}
      \An{n}{1}{k} & \An{n}{13}{k} & \An{n}{9}{k} & \An{n}{5}{k} \\
      \Dn{n}{16}{k}& \Dn{n}{12}{k}& \Dn{n}{8}{k}& \Dn{n}{4}{k}\\
      \Cn{n}{11}{k}& \Cn{n}{7}{k}& \Cn{n}{3}{k}& \Cn{n}{15}{k}\\
      \Bn{n}{6}{k}& \Bn{n}{2}{k}& \Bn{n}{14}{k}& \Bn{n}{10}{k}
    \end{array}
  \right)
  \nonumber\\
  \fl = 
  \left(
    \begin{array}{c@{\enspace}c@{\enspace}c@{\enspace}c}
      P_{00} & P_{10} & P_{20} & P_{30} \\
      \phi_k P_{01} & \phi_k P_{11} & \phi_k P_{21} & \phi_k P_{31} \\
      \phi_k^2 P_{02} & \phi_k^2 P_{12} & \phi_k^2 P_{22} & \phi_k^2 P_{32} \\
      \phi_k^3 P_{03} & \phi_k^3 P_{13} & \phi_k^3 P_{23} & \phi_k^3 P_{33} 
    \end{array}
  \right)
  \left(
    \begin{array}{c@{\enspace}c@{\enspace}c@{\enspace}c}
      \An{n-1}{1}{k} & \An{n-1}{13}{k} & \An{n-1}{9}{k} & \An{n-1}{5}{k} \\
      \Dn{n-1}{16}{k}& \Dn{n-1}{12}{k}& \Dn{n-1}{8}{k}& \Dn{n-1}{4}{k}\\
      \Cn{n-1}{11}{k}& \Cn{n-1}{7}{k}& \Cn{n-1}{3}{k}& \Cn{n-1}{15}{k}\\
      \Bn{n-1}{6}{k}& \Bn{n-1}{2}{k}& \Bn{n-1}{14}{k}& \Bn{n-1}{10}{k}
    \end{array}
  \right),
  \nonumber\\
  \fl = 
  \left(
    \begin{array}{c@{\enspace}c@{\enspace}c@{\enspace}c}
      P_{00} & P_{10} & P_{20} & P_{30} \\
      \phi_k P_{01} & \phi_k P_{11} & \phi_k P_{21} & \phi_k P_{31} \\
      \phi_k^2 P_{02} & \phi_k^2 P_{12} & \phi_k^2 P_{22} & \phi_k^2 P_{32} \\
      \phi_k^3 P_{03} & \phi_k^3 P_{13} & \phi_k^3 P_{23} & \phi_k^3 P_{33} 
    \end{array}
  \right)^{n-1}
  \left(
    \begin{array}{c@{\enspace}c@{\enspace}c@{\enspace}c}
      \An{1}{1}{k} & \An{1}{13}{k} & \An{1}{9}{k} & \An{1}{5}{k} \\
      \Dn{1}{16}{k}& \Dn{1}{12}{k}& \Dn{1}{8}{k}& \Dn{1}{4}{k} \\
      \Cn{1}{11}{k}& \Cn{1}{7}{k}& \Cn{1}{3}{k}& \Cn{1}{15}{k}\\
      \Bn{1}{6}{k}& \Bn{1}{2}{k}& \Bn{1}{14}{k}& \Bn{1}{10}{k}
    \end{array}
  \right),
  \nonumber\\
  \fl = 
  \left(
    \begin{array}{c@{\enspace}c@{\enspace}c@{\enspace}c}
      P_{00} & P_{10} & P_{20} & P_{30} \\
      \phi_k P_{01} & \phi_k P_{11} & \phi_k P_{21} & \phi_k P_{31} \\
      \phi_k^2 P_{02} & \phi_k^2 P_{12} & \phi_k^2 P_{22} & \phi_k^2 P_{32} \\
      \phi_k^3 P_{03} & \phi_k^3 P_{13} & \phi_k^3 P_{23} & \phi_k^3 P_{33} 
    \end{array}
  \right)^{n-1}
  \left(
    \begin{array}{c@{\enspace}c@{\enspace}c@{\enspace}c}
      P_{00} & \phi_k P_{10} & \phi_k^2 P_{20} & \phi_k^3 P_{30}\\
      \phi_k P_{01} & \phi_k^2 P_{11} & \phi_k^3 P_{21} & P_{31}\\
      \phi_k^2 P_{02} & \phi_k^3 P_{12} & P_{22} & \phi_k P_{32}\\
      \phi_k^3 P_{03} & P_{13} & \phi_k P_{23} & \phi_k^2 P_{33} 
    \end{array}
  \right),\nonumber\\
  = 
  \left(
    \begin{array}{c@{\enspace}c@{\enspace}c@{\enspace}c}
      P_{00} & P_{10} & P_{20} & P_{30} \\
      \phi_k P_{01} & \phi_k P_{11} & \phi_k P_{21} & \phi_k P_{31}
      \\ 
      \phi_k^2 P_{02} & \phi_k^2 P_{12} & \phi_k^2 P_{22} & \phi_k^2
      P_{32} \\ 
      \phi_k^3 P_{03} & \phi_k^3 P_{13} & \phi_k^3 P_{23} & 
      \phi_k^3 P_{33} 
    \end{array}
  \right)^{n}
  \left(
    \begin{array}{c@{\enspace}c@{\enspace}c@{\enspace}c}
      1 & 0 & 0 & 0 \\
      0 & \phi_k & 0 & 0 \\
      0 & 0 & \phi_k^2 & 0 \\
      0 & 0 & 0 & \phi_k^3
    \end{array}
  \right)\,,
  \label{abcdsumrecursion}
\end{eqnarray}
provided 
\begin{equation}
  \phi_k^4 = 1 \Leftrightarrow
  \phi_k = \exp(\ii k\pi/2), \quad k = 0,1,2,3\,,
\end{equation}

We have the following identities,
\begin{eqnarray*}
  2[\an{n}{1} - \an{n}{11}] = \An{n}{1}{1} + \An{n}{1}{3}\,,\\
  2[\an{n}{5} - \an{n}{15}] = \An{n}{5}{1} + \An{n}{5}{3}\,,\\
  2[\an{n}{9} -\an{n}{3}] = \An{n}{9}{1} + \An{n}{9}{3} \,,\\
  2[\an{n}{13} -\an{n}{7}] = \An{n}{13}{1} + \An{n}{13}{3} \,,\\
  2[\bn{n}{2} - \bn{n}{12}] =  \Bn{n}{2}{1}+ \Bn{n}{2}{3} \,,\\
  2[\bn{n}{6} - \bn{n}{16}] = \Bn{n}{6}{1} + \Bn{n}{6}{3}\,,\\
  2[\bn{n}{10} - \bn{n}{4}] = \Bn{n}{10}{1} + \Bn{n}{10}{3}\,,\\
  2[\bn{n}{14} - \bn{n}{8}] = \Bn{n}{14}{1} + \Bn{n}{14}{3} \,,\\
  2[\cn{n}{3} - \cn{n}{9}] = \Cn{n}{3}{1} + \Cn{n}{3}{3}\,,\\
  2[\cn{n}{7}  - \cn{n}{13}] = \Cn{n}{7}{1} + \Cn{n}{7}{3}\,,\\
  2[\cn{n}{11} - \cn{n}{1}] = \Cn{n}{11}{1} + \Cn{n}{11}{3}\,,\\
  2[\cn{n}{15} - \cn{n}{5}] = \Cn{n}{15}{1} + \Cn{n}{15}{3}\,,\\
  2[\dn{n}{4} -\dn{n}{10}] = \Dn{n}{4}{1} + \Dn{n}{4}{3}\,,\\
  2[\dn{n}{8} - \dn{n}{14}] = \Dn{n}{8}{1} + \Dn{n}{8}{3}\,,\\
  2[\dn{n}{12} - \dn{n}{2}] = \Dn{n}{12}{1} + \Dn{n}{12}{3}\,,\\
  2[\dn{n}{16} - \dn{n}{6}] = \Dn{n}{16}{1} + \Dn{n}{16}{3}\,,
\end{eqnarray*}
which, using equation (\ref{abcdsumrecursion}) yield the velocity
auto-correlation (\ref{v0vn2smasqexp}).

\end{appendix}

\end{document}